\pdfoutput=1
\documentclass[manuscript,nonacm]{acmart}
\AtBeginDocument{%
  }

\usepackage{subcaption}
\usepackage{float}
\usepackage{tcolorbox}
\usepackage{enumitem}
\tcbuselibrary{breakable, skins}
\newtcolorbox{promptbox}[1][]{
  colback=gray!3,
  colframe=gray!60,
  fonttitle=\bfseries\footnotesize,
  title={#1},
  breakable,
  enhanced,
  left=3pt, right=3pt, top=3pt, bottom=3pt,
  fontupper=\scriptsize,
  before upper={\setlength{\parskip}{4pt}\setlength{\parindent}{0pt}}
}

\newcommand{\ie}{\textit{i}.\textit{e}.}
\newcommand{\eg}{\textit{e}.\textit{g}.}

\newcommand{\systemname}{Brief2Design}

\newcommand{\approvalnumber}{CAE-2024-14}

\newcommand{\mention}[2]{\textit{``#2''} (#1)}

\begin{document}

\title[Brief2Design]{Brief2Design: A Multi-phased, Compositional Approach to Prompt-based Graphic Design}

\author{Kotaro Kikuchi}
\affiliation{%
  \institution{CyberAgent}
  \city{Tokyo} %
  \country{Japan}
}
\email{kikuchi_kotaro_xa@cyberagent.co.jp}

\author{Nami Ogawa}
\affiliation{%
  \institution{CyberAgent}
  \city{Tokyo} %
  \country{Japan}
}

\renewcommand{\shortauthors}{Kikuchi and Ogawa}

\begin{abstract}
Professional designers work from client briefs that specify goals and constraints but often lack concrete design details. Translating these abstract requirements into visual designs poses a central challenge, yet existing tools address specific aspects or induce fixation through complete outputs. Through interviews with six professional designers, we identified how designers address this challenge: first structuring ambiguous requirements, then exploring individual elements, and finally recombining alternatives. We developed Brief2Design, supporting this workflow through requirement extraction and recommendation, element-level exploration for objects, backgrounds, text, typography, and composition, and flexible recombination of selected elements. A within-subjects study with twelve designers compared Brief2Design against a conversational baseline. The structured approach increased prompt diversity and received high ratings for requirement extraction and recommendation, but required longer generation time and achieved comparable image diversity. These findings reveal that structured workflows benefit requirement clarification at the cost of efficiency, informing design trade-offs for AI-assisted graphic design tools.

\end{abstract}



\begin{teaserfigure}
  \centering
  \includegraphics[width=\textwidth]{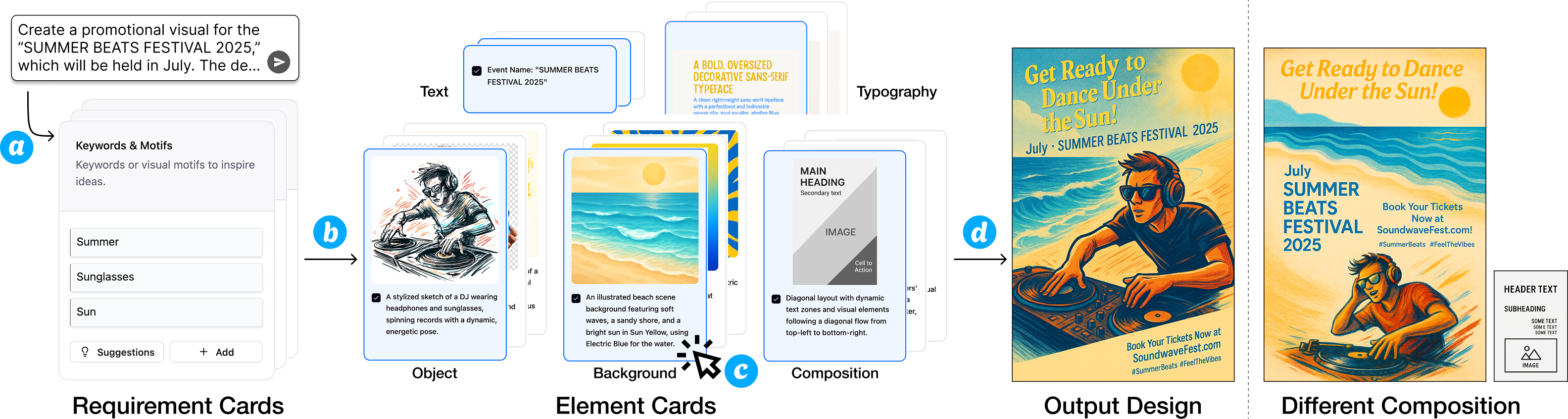}
  \caption{Brief2Design is an end-to-end system that supports designers in creating graphic designs from design briefs. The system first (a) structures the input text into requirement cards and (b) generates type-specific element cards. After designers (c) select the element cards, the system (d) integrates the selected elements into a single design. The designers can easily explore design variations by recombining element cards, such as changing the composition cards.}
  \label{fig:teaser}
\end{teaserfigure}

\maketitle

\section{Introduction}

In professional graphic design practice, designers typically work from client requests that specify project goals, target audiences, and constraints. These requests, which we refer to as design briefs in this paper, may take various forms from formal written documents to informal conversations~\cite{JonesDesign2012,ReadFunctions2012}. Translating these high-level requirements into concrete design decisions, such as selecting appropriate visual content, crafting effective copy, determining typography, and arranging layouts, represents a central challenge of the design process. Furthermore, exploring the design space by considering multiple alternatives is known to lead to better and more diverse design outcomes~\cite{Dow2010-mb}, but this iterative exploration requires significant expertise, time, and effort.

Recent advances in generative AI, particularly text-to-image (T2I) models and large language models (LLMs), have shown great potential in supporting creative tasks~\cite{OpenAIIntroducing2025,RaisinghaniIntroducing2025}, raising expectations for their application in graphic design. However, existing approaches have been developed for different scenarios and thus do not fully support the brief-to-design workflow central to professional practice.
The dominant conversational approach uses systems like ChatGPT with T2I integration, assuming users start from scratch and iteratively generate and refine complete images through dialogue. While this approach is accessible, it proves time-consuming and misaligned with the non-linear nature of design processes such as remixing ideas~\cite{Zhou2024-fd}. Moreover, from a cognitive perspective, early exposure to high-quality generated images can induce design fixation~\cite{Wadinambiarachchi2024-uz}, where designers become anchored to AI-presented solutions and struggle to explore alternatives independently. Effective design processes require systematic exploration of the design space before committing to specific directions~\cite{Suh2024-de}.
Creativity support tools have also emerged across various design contexts, such as graphic design ideation~\cite{Choi2024-zc}, infographics~\cite{ZhouEpigraphics2024}, UI/UX design~\cite{ShokrizadehDancing2025}, and social media content~\cite{Liu2025-pm}. While these systems address aspects of the design process, none specifically supports the brief-to-design workflow where designers must first analyze abstract client requests before exploring concrete design elements.

To understand the practices and challenges professional designers face in this workflow, we conducted a formative study with six professional designers with 10-28 years of experience ($M = 17.7$, $SD = 6.7$) specializing in various graphic design domains including banner ads, game interfaces, posters, and logos.
Through semi-structured interviews, we observed that designers consistently start by analyzing the given conditions and requirements, which are often abstract or insufficient in detail. They systematically complement missing requirements through additional inquiries to clients, comparisons with past projects, and their own professional judgment based on context and experience. After establishing this grounding, they proceed to more concrete design considerations, exploring visual content, crafting copy, selecting typography, and determining layouts, while maintaining an iterative process that explores multiple alternatives and flexibly recombines promising elements from different directions.
Based on these observations, we formulated three design goals for AI-assisted graphic design systems: (1) support structured analysis and exploration of design requirements, (2) enable element-level exploration with visual feedback, and (3) facilitate flexible recombination across alternatives.

To address these goals, we propose \textit{Brief2Design}, a T2I-based graphic design system specifically designed to support the brief-to-design workflow in professional practice. Unlike conversational systems that iteratively generate complete images or creativity support tools that address specific design aspects, \textit{Brief2Design} begins by analyzing design briefs to extract explicit requirements and identify gaps that need to be filled.
The system supports a structured three-step workflow: (1) \textit{requirement-level grounding} that analyzes briefs and helps designers explore missing requirements through LLM-based requirement extraction and gap identification, (2) \textit{element-level exploration} that provides visual previews for specific design aspects such as copy, visual content, and style, enabling designers to explore alternatives systematically, and (3) \textit{integrated design generation} that allows flexible recombination of curated elements into final compositions through a T2I model with compositional control.
We carefully designed LLM-based components for each step to bridge the gap between abstract client requests and concrete design outcomes, while preserving designers' control over creative decisions.

We evaluated \textit{Brief2Design} through a within-subjects study ($N = 12$) comparing it against a conversational baseline using the same T2I model.
The structured workflow significantly increased prompt diversity and received high ratings for requirement extraction features, demonstrating the value of explicit brief clarification and element-level exploration.
However, it also led to longer time per generation and lower task completion rates under time constraints, while revealing evaluation challenges for context-dependent elements such as typography and composition.
These findings highlight trade-offs between structured exploration and workflow efficiency in AI-assisted graphic design.

This paper makes the following contributions:
\begin{itemize}
    \item \textit{Brief2Design}, a T2I-based graphic design system that supports the brief-to-design workflow through structured requirement analysis, element-level exploration, and flexible recombination.
    \item Empirical findings on structured workflows revealing both benefits in requirement clarity and costs in workflow efficiency, with implications for balancing scaffolding and flexibility in AI-assisted design tools.
\end{itemize}

\section{Related Work}

\subsection{Text-to-Image Models and Design-Oriented Generation}
T2I generative models, typically based on diffusion architectures, synthesize images from natural language prompts by learning text-image correspondences at scale. Recent models can follow complex instructions and produce legible text and layout-sensitive compositions, which has broadened their use in early-stage design exploration.
However, producing reliable outcomes that are in line with the user's intent depends strongly on prompt craft. Prior work highlights both the importance and the difficulty of prompt authoring: a study exploring how non-AI-experts designing prompts found that they tend to prompt opportunistically rather than systematically~\cite{Zamfirescu-Pereira2023-yt}, while Liu and Chilton~\cite{Liu2022-bw} offer design guidelines for T2I prompt engineering through a series of experiments. These findings suggest that systems that require users to naively iterate on textual prompts may not be ideal, especially for non-AI experts.

To address these challenges, systems have explored structured support for T2I prompt design.
Recent studies have developed systems to help users design prompts effectively and interactively. 
For example, Promptify~\cite{Brade2023-gb} facilitates prompt exploration and refinement by offering automatic subject and style suggestions and supporting image management through similarity clustering.
PromptCharm also supports users' prompt refinement by automatic prompt optimization and a multimodal feedback loop via attention visualization, keyword attention adjustment, and image inpainting~\cite{Wang2024-ym}.
Another line of studies proposes more structured interfaces: PromptMagician~\cite{Feng2024-em} is a visual analysis system that recommends prompt keywords, helping users efficiently explore generated images and refine input prompts and hyperparameters. DreamSheets~\cite{Almeda2024-mi} is a spreadsheet-based interface that allows users to author workflows to explore a vast design space for T2I models. Together, these works move from free-form chat toward interfaces that externalize and organize the prompt space.

A complementary line of studies targets design-oriented generation.
Early research in this area focused on individual design elements such as typography, layout, and color palette, employing specialized models for each subtask~\cite{ZouFragment2025}.
More recently, systems have shifted toward integrated approaches that generate complete designs.
ART~\cite{PuART2025} enables direct generation of variable multi-layer transparent images based on text prompts and anonymous region layouts, facilitating layer-level isolation and editing.
COLE~\cite{JiaCOLE2024} and OpenCOLE~\cite{InoueOpenCOLE2024} introduce hierarchical generation pipelines that decompose user intent into multi-layered layout planning, followed by separate generation of background, objects, and typography, enabling flexible layer-wise editing.
POSTA~\cite{ChenPOSTA2025} similarly adopts a modular pipeline with enhanced artistic text stylization.
BannerAgency~\cite{WangBannerAgency2025} takes a training-free agentic approach, supporting diverse sizes, genres, and languages.
These approaches illustrate pathways for aligning T2I with graphic design needs and requirements.

While these systems advance T2I prompt authoring and design-oriented generation, they assume users already have clear design directions or specific content determined.
In contrast, professional design workflows often begin with abstract client briefs that require systematic analysis and iterative exploration of underspecified requirements before moving to concrete design production.
We contribute an end-to-end workflow that structures this process from brief analysis through element-level exploration to flexible composition.

\subsection{Creativity Support Tools for Design}
Early research developed creativity support tools (CSTs) that approach graphic design by decomposing it into individual elements and optimizing them based on codified design principles or data-driven aesthetic preferences~\cite{O-Donovan2015-qs, Guo2021-ky, YangAutomatic2016, KongAesthetics2023,Zhao2020-rz}.
DesignScape~\cite{O-Donovan2015-qs} improves layouts by suggesting refinements to element position, scale, and alignment based on computational design principles.
Vinci~\cite{Guo2021-ky} automatically generates advertising posters from user-specified product images and taglines, integrating online editing-feedback that allows users to update results based on their design preferences.
These systems demonstrated that decomposing design into manipulable facets (\eg, layout, typography) and applying computational optimization can scaffold novices and accelerate iteration for experts.

With the advancement of large language models (LLMs) and their agentic use, addressing domain-specific design challenges beyond aesthetic preferences has become more practical, enabling systems that target graphic design~\cite{Choi2024-zc}, video editing~\cite{Wang2024-xr}, infographics~\cite{ZhouEpigraphics2024}, UI/UX~\cite{ShokrizadehDancing2025}, and social media content~\cite{Liu2025-pm}.
CreativeConnect~\cite{Choi2024-zc} supports graphic designers' ideation phases by recombining reference images through keyword extraction, recommendation, and sketch generation.
Epigraphics~\cite{ZhouEpigraphics2024} treats the designer's message as a first-class object, using it to guide infographic asset creation, editing, and syncing through recommendations for visualizations, graphics, and color palettes.
UIDEC~\cite{ShokrizadehDancing2025} supports UI/UX ideation under constraints by allowing designers to specify project details (\eg, purpose, target audience, industry, design styles) and generating diverse examples that adhere to these constraints.
Influencer~\cite{Liu2025-pm} empowers design novices to create promotional posts through multidimensional recommendations for images and captions, context-aware exploration, and flexible conceptual fusion.
DesignManager~\cite{YouDesignManager2025} integrates multiple AI design tools into creative workflows through node-based process visualization and agentic context management.

While individual systems incorporate relevant features (such as message-driven design guidance~\cite{ZhouEpigraphics2024}, constraint-based exploration of UI designs~\cite{ShokrizadehDancing2025}, or concept fusion~\cite{Liu2025-pm}), no system integrates the complete brief-to-design workflow that professional practice requires.
Our system addresses this gap by combining systematic brief analysis and requirement structuring, requirement-based exploration at the element level for recombination, and controlled integration of these elements across multiple design aspects (objects, backgrounds, text, typography, composition) into final designs.
Through this system, we investigate how structured workflows affect the design exploration process and associated trade-offs.

\subsection{Design Principles and Cognitive Challenges in Human-AI Co-Creation}
Researchers have developed methodological foundations for structuring human-AI co-creation processes~\cite{Wu2022-zt, Cao2025-sa, Zhou2024-fd, GmeinerIntent2025}.
Wu et al.~\cite{Wu2022-zt} demonstrated that decomposing complex generation tasks into sequential steps with visible intermediate artifacts improves transparency and controllability, enabling users to inspect and modify outputs at each stage.
Cao et al.~\cite{Cao2025-sa} proposed organizing creative content through compositional structures (spatial, narrative, congruent, temporal) that synchronize different aspects of a project, allowing changes in one structure to propagate to others while maintaining consistency.
Zhou et al.~\cite{Zhou2024-fd} emphasized that creative processes are inherently non-linear, advocating for AI systems that support iterative refinement, clarification through dialogue, and remixing of alternative solutions rather than linear command execution.
These works establish design principles for human-AI co-creation workflows that balance structure with flexibility.

From a cognitive perspective on human-AI co-creation, creators often face the challenge of design fixation, where they become unconsciously anchored to existing examples or early ideas~\cite{JanssonDesign1991,YoumansDesign2014}.
Effective exploration of the design space has been shown to mitigate this issue~\cite{Suh2024-de,Tao2025-zu,Jeon2021-wu}.
For instance, Luminate~\cite{Suh2024-de} structures the design space by generating multiple conceptual dimensions (\eg, tone, perspective) and populating them with diverse options, enabling writers to systematically explore alternatives before committing to specific directions.
Also in the context of generative AI, empirical studies reveal that early exposure to AI-generated complete outputs can induce design fixation and reduce ideation fluency and diversity~\cite{Wadinambiarachchi2024-uz}, as highly polished AI outputs anchor users to presented solutions and inhibit independent exploration.

Following these design principles, our system decomposes the brief-to-design workflow into three steps with editable intermediate artifacts (requirement cards and element cards), organizes design content through compositional structures that synchronize requirements and elements, and supports non-linear exploration by allowing users to iterate within and across steps.
Additionally, to address the cognitive challenges of design fixation, the system facilitates design space exploration through dedicated interfaces for requirements and elements.
By guiding users through explicit stages of requirement, element, and integration, users can consider alternatives at each level before generating complete compositions, thereby avoiding early exposure to polished outputs that might anchor their thinking.

\section{Formative Study}
We conducted interviews with designers to understand how professional designers work from client briefs in practice, identifying their workflows and the challenges they face in the brief-to-design process. This chapter provides an overview of the study and its key findings.

\subsection{Participants and Procedure}
We recruited six designers (D1-6; 4 men and 2 women, age: $M = 40.3$, $SD = 5.5$) from our company. Some designers specialized in banner ads (D1, D3) or game user interfaces (D2), while others (D4, D5, D6) had experience in various design fields, including posters, catalogs, and logos. The participants had professional experience ranging from 10 to 28 years ($M = 17.7$, $SD = 6.7$).

The study was conducted as a 60-minute semi-structured interview via online conferencing. The interview focused on five main topics: (1) how design requests are received and the requirements definition process, (2) methods for idea generation and variation design, (3) procedures for determining visual elements, and (4) feedback and revision workflows. Participants in this study participated as part of their work duties, and no additional expenses were incurred.
The obtained transcripts and materials were analyzed to extract common behavioral patterns and challenges. This formative study and the later user evaluation were approved by the internal ethics review board of our company (Approval No. \approvalnumber).

\subsection{Findings}

Through our interviews, we identified four key aspects of professional design practice: understanding and supplementing ambiguous requests, defining and structuring design concepts, exploring design elements individually, and combining and integrating multiple elements into final designs. While these practices demonstrate designers' expertise and adaptability, they also reveal time-consuming processes that rely heavily on domain knowledge and iterative refinement. These observations inform the design of AI support systems that can assist designers while preserving their creative control.

\subsubsection{Understanding and Supplementing Ambiguous Requests}
The clarity and specificity of design requests varied significantly. While some requests came with detailed specifications or rough drafts (D5), others only conveyed the general context and target audience (D1, D3). D3 noted, \textit{``Sometimes there are very few shared details, and by asking questions, we can uncover the request's content. Experienced salespeople often provide detailed information about what we need.''} 
In cases of insufficient information, designers supplemented the content through additional inquiries to the requester (D3, D5), comparisons with past or similar projects (D1, D3, D5), and their own experience or external resources (D1, D4, D6).
A common approach was observed where designers distinguished between constraints and areas of discretion within the provided information to organize their production strategy. D1 mentioned, \textit{``To summarize, it involves adhering to the client's regulations, incorporating targeting information from sales consultants, and reflecting any appeals they want to try out. Beyond that, the production is quite flexible.''}
However, this requirement supplementation process requires time and cognitive effort, particularly for designers with less experience in specific domains. Analyzing ambiguous briefs to identify what is specified and what is missing, and systematically exploring possibilities for incomplete requirements, represents a substantial burden in the early stages of design work.

\subsubsection{Defining and Structuring Design Concepts}
During the ideation phase, designers navigated between the core intent to be conveyed and various perspectives on how to express it.
D6 often began by writing down keywords in a notebook to explore ideas: \textit{``First, I write down the overview of the logo to keep it consistent, then I jot down keywords and expand them like a mind map. If I find an interesting direction, I might think, `Let's combine this and that.' I often organize information in writing at this stage, avoiding drawing as much as possible until I've sorted out the core words.''} D6 also uses ChatGPT to assist in verbalizing difficult concepts: \textit{``Tell the content received in the briefing and ask, 'Please give me as many keywords as possible.' For the keywords that come up, I ask, 'What does this mean?' or 'Please provide related keywords,' to dig deeper. Sometimes I ask for visual ideas, but that often doesn't work well, so I mainly use it to assist in concept building.''}
D2, who specializes in game user interfaces, described their approach as follows: \textit{``When deciding on the design direction, I first look at the rough game design document created by the director or producer during game development, which outlines the general direction of the game. This document includes information such as the target audience and persona settings, which I use to pick up relevant points, discuss them, verbalize them, and make decisions.''}
We observed that designers took time to organize the direction and concept before starting concrete production, defining the exploratory domain of design.
This conceptualization process involves trial and error in articulating abstract ideas, which can be time-consuming and requires designers to explore multiple directions before converging on a clear direction.

\subsubsection{Element-wise Exploration}
At each stage of production, designers formed the overall impression while individually considering elements such as layout, typography, color, and visual materials.
For layout, D4 started by placing all the information and \mention{P4}{adjusting the layout while organizing the hierarchy of importance among elements.}.
Some designers decided on typography early: \mention{D2}{The typography largely determines the overall impression of the design $\dots$ so I try several variations from the available fonts.}.
For visual materials, designers used both stock photo sites and generative AI, repeatedly creating and testing alternatives. As one participant described, \mention{D4}{I generate several patterns, download the ones that seem promising, place them, and if they don't quite fit, I create new ones.}.
Through these iterative explorations of individual elements, designers gradually refined the composition as a whole, achieving coherence between the details and the overall impression.
This iterative refinement at the element level requires repeated generation, evaluation, and adjustment cycles.

\subsubsection{Combination and Integration of Elements}
Beyond examining individual elements, designers also engaged in combining and integrating multiple ideas or components to achieve more refined outcomes. This process was characterized by the selective synthesis of valuable aspects from different directions, leading to the emergence of new possibilities within the design.
For instance, D1 viewed advertising design as \textit{``a multiplication of copy and visuals''}, exploring stored copies and images while searching for combinations that seemed interesting — \textit{``it might be fun to pair this with that.''} D2's UI design process similarly involved integrating ideas proposed by multiple designers. Rather than simply selecting one version, they often \textit{``took the good parts from both,''} such as deciding, \textit{``the font from this one works better, so let's use it here.''} In this way, multiple perspectives were merged into a single cohesive design.
In D4's logo design, after the first proposal, clients occasionally requested to \textit{``combine this and that to make a new one,''} prompting designers to reassemble existing drafts into new forms. Likewise, D6 prepared multiple preliminary versions and, when two designs showed potential, \textit{``fused them and further refined the result.''}
Through such practices of combination and integration, designers did not merely choose among alternatives but constructed new creative value by synthesizing multiple perspectives, making this process a central part of design production.
However, this recombination process can be challenging when working with complete images, as it requires manual editing or regeneration from scratch to extract and reassemble specific elements.

\section{Design Goals}

Our formative study revealed that professional designers engage in four key practices: supplementing ambiguous requirements, structuring design concepts, exploring individual elements, and recombining elements across alternatives. While these practices demonstrate expertise, they also involve time-consuming processes that rely heavily on domain knowledge. Specifically, designers face challenges in (1) analyzing ambiguous briefs to identify specified requirements and explore missing ones, (2) articulating abstract concepts through trial and error to define design directions, (3) iteratively refining elements through repeated generation and evaluation cycles, and (4) manually extracting and reassembling specific elements when working with complete images.

Based on these observations, we formulate three design goals for AI-assisted graphic design systems that support professional designers working from client briefs:

\begin{itemize}
    \item \textbf{DG1: Support Structured Analysis and Exploration of Design Requirements.} Systems should help designers transform ambiguous briefs into structured, explorable representations that make explicit what is specified and what remains to be defined, facilitating systematic exploration of missing requirements.
    \item \textbf{DG2: Enable Element-Level Exploration with Visual Feedback.} Systems should allow designers to explore individual design aspects separately with immediate visual feedback, reducing the need for repeated full-composition generation cycles while maintaining overview of how elements contribute to the whole.
    \item \textbf{DG3: Facilitate Flexible Recombination Across Alternatives.} Systems should enable designers to freely recombine promising elements from different explorations into new compositions, supporting the synthesis of multiple design directions without requiring manual extraction or regeneration.
\end{itemize}

\section{\systemname{}}

\subsection{Overview}

To realize these design goals, we propose \systemname{}, a T2I-based graphic design system that instantiates these principles through a three-step workflow. The system addresses DG1 by structuring free-form briefs into editable requirement cards and recommending missing requirements, DG2 by generating preview images for individual design elements before final composition, and DG3 by maintaining elements as separate entities that can be freely recombined across different compositions.

The system consists of three main steps: (1) Requirement Extraction and Recommendation, (2) Element-level Design Recommendation and Selection, and (3) Final Design Generation via Element Integration, as illustrated in Figure~\ref{fig:teaser}. Brief2Design centers around a collaborative design process where ``AI suggests, and humans decide.'' At each step, users can review AI recommendations and adjust requirements and elements according to their intentions. This approach enables a user experience that reflects human intentions sequentially, rather than generating a complete design first and subsequently refining it through repeated feedback cycles, as is common with conventional generation tools. The expected user experience flow is as follows:
\begin{enumerate}
    \item When a user inputs requirements in natural language, the AI structures them into editable requirement cards.
    \item Next, the AI recommends candidate design elements based on the requirements, allowing users to select or edit them while previewing images.
    \item Finally, the final design is generated based on the selected elements, and users can explore other possibilities by comparing with past results or trying different combinations of elements.
\end{enumerate}

Figures~\ref{fig:screen_step1}--\ref{fig:screen_step3} present the stage-specific interfaces of \systemname{}.

\subsection{Step 1: Requirement Extraction and Recommendation}

\begin{figure}[t]
  \centering
  \includegraphics[width=\columnwidth]{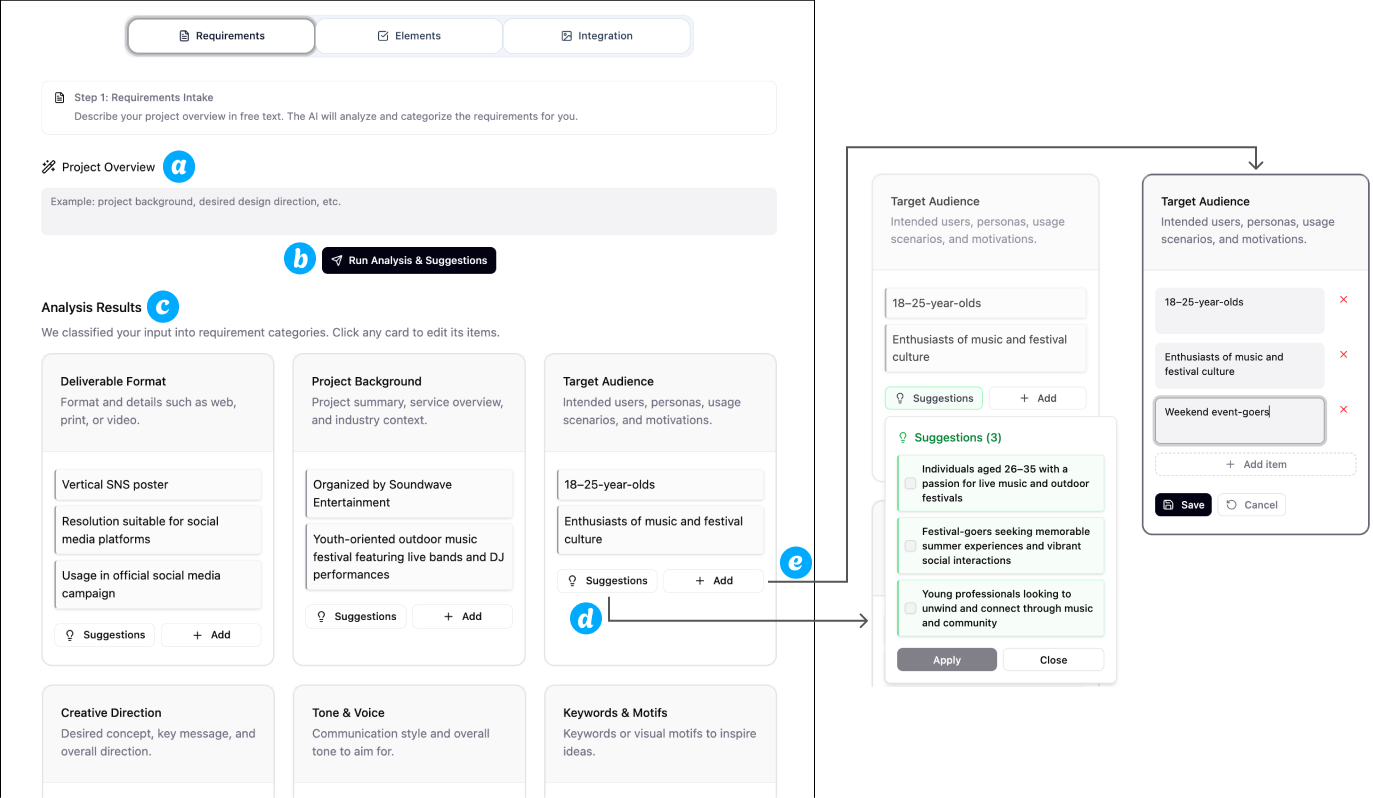}
  \caption{Step 1 interface: Structuring free-form briefs into requirement cards. Users input their design brief~(a) and trigger requirement extraction~(b), which categorizes the input into eight requirement cards~(c). Users can then request AI-generated recommendations~(d) for each category or manually add custom entries~(e).}
  \label{fig:screen_step1}
\end{figure}

This step focuses on structuring and visualizing the user's design requirements, which are initially provided in free-form text. The goal is to facilitate the user's consideration of their design intent by transforming ambiguous natural language descriptions into clearly organized requirement cards (Figure~\ref{fig:screen_step1}).

There are two modules involved in this step: \emph{RequirementExtractor} and \emph{RequirementRecommender}. RequirementExtractor processes the user's input text to extract and categorize key design requirements into eight predefined fields: ``Deliverable Format,'' ``Business Context,'' ``Target Audience,'' ``Creative Direction,'' ``Tone and Manner,'' ``Keywords and Motifs,'' ``Design Specifications,'' and ``Restrictions.'' Each field has a corresponding card and can contain multiple entries. RequirementRecommender analyzes the existing requirements and suggests $N$ candidate entries for the target fields. The candidates are designed to be distinct from the existing ones and to align with other requirement fields.

The user can interact with each requirement card to add new entries, edit existing ones, or delete them. Additionally, users can request AI-generated recommendations for new entries and choose to accept or reject them. This interactive process enables users to refine and expand their design requirements in collaboration with AI.

Once the user is satisfied with the set of requirement cards, they can proceed to the next step. This process is inherently iterative, allowing users to revisit and refine their requirements at any time. Rather than merely functioning as a preprocessing stage for generation, this step serves as an interactive environment for externalizing and organizing design intent. By visualizing requirements as editable structures, the system encourages reflective thinking and enables users to evolve their design briefs through interaction with the AI collaboratively.

\subsection{Step 2: Element-level Design Recommendation and Selection}

\begin{figure}[t]
  \centering
  \includegraphics[width=\columnwidth]{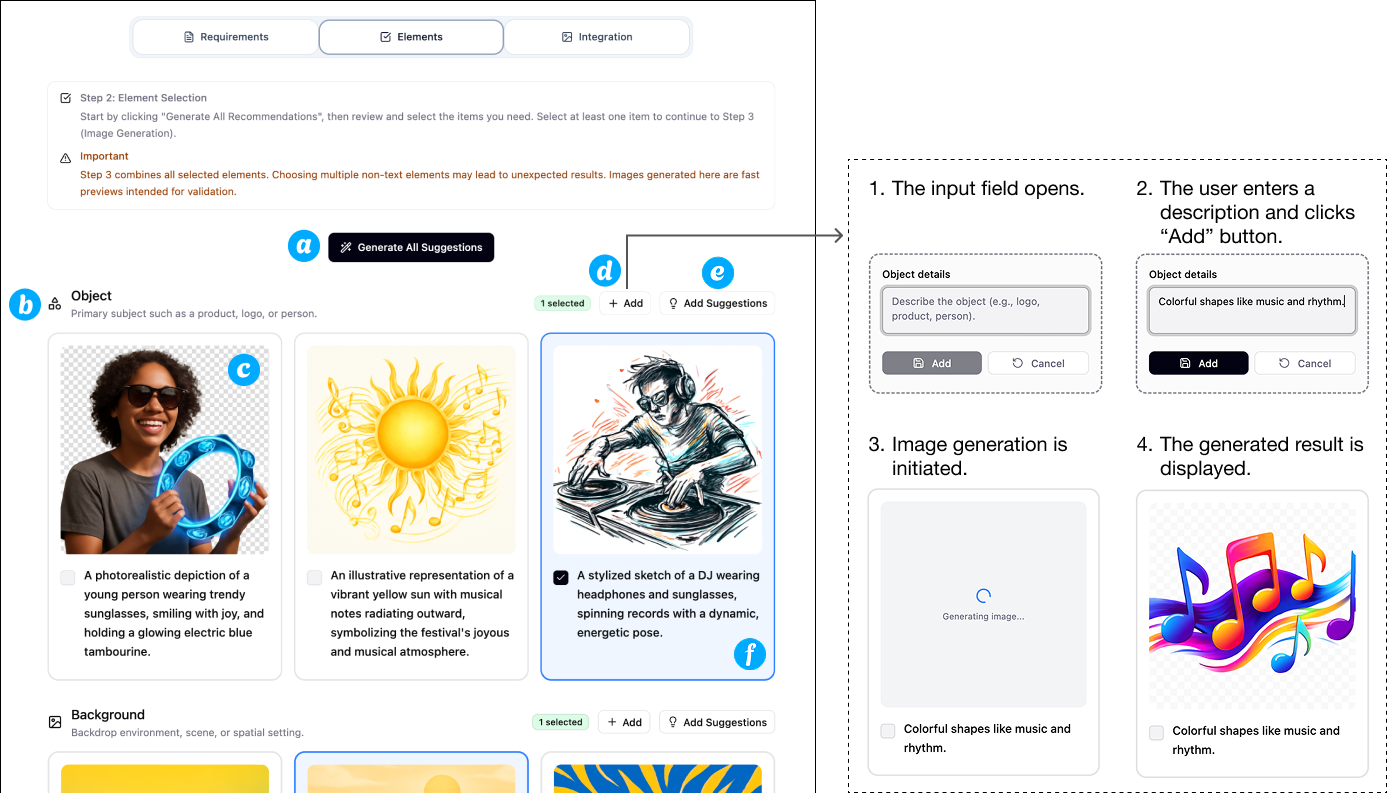}
  \caption{Step 2 interface: Generating type-specific element cards with preview images. Users initiate element generation~(a), and the system displays element cards for five categories: Objects, Backgrounds, Text, Typography, and Composition~(b). Cards for visual elements include preview images~(c), enabling users to visually evaluate each element. At the category level, users can manually add new candidates~(d) or request additional AI recommendations~(e). Users can then select their preferred elements~(f) for integration in the next step.}
  \label{fig:screen_step2}
\end{figure}

This step focuses on the process where AI recommends design elements based on the structured requirement cards, and users form element-level design configurations by editing and adjusting them (Figure~\ref{fig:screen_step2}). The goal is to expand the abstract design intent described in the requirements into visually verifiable concrete elements, such as \emph{Objects}, \emph{Backgrounds}, \emph{Text}, \emph{Typography}, and \emph{Composition}, enabling users to explore design directions through collaborative trials with AI.

This step involves two main modules: \emph{ElementRecommender} and \emph{ElementVisualizer}. ElementRecommender takes the organized requirement cards from the previous step as input, along with guidelines for the target element types and context information about already recommended elements of the same type. Based on this, it uses LLMs to generate $N$ \emph{rough prompts} for each element type. Each candidate is designed to explore different design directions while maintaining consistency with the requirements. ElementVisualizer then takes these rough prompts and is responsible for refining them and generating preview images. It applies different enhancement guidelines based on the element type, transforming the rough prompts into \emph{enhanced prompts} optimized for image generation. For example, for \emph{Objects}, it specifies details such as texture and lighting conditions, applying different descriptive strategies based on subtypes like ``Commercial/Professional,'' ``Natural/Character,'' and ``Artistic/Design.'' For \emph{Composition}, it uses enhancement templates that quantitatively specify element placement, converting layout descriptions into specific values like ``top 25\%'' or ``center 60\%.'' Through these type-specific enhancement templates, ElementVisualizer converts ambiguous text descriptions into high-precision structured prompts and generates corresponding preview images. The generated enhanced prompts are also reused in the subsequent integration generation step.

\begin{figure}[t]
  \centering
  \includegraphics[width=\columnwidth]{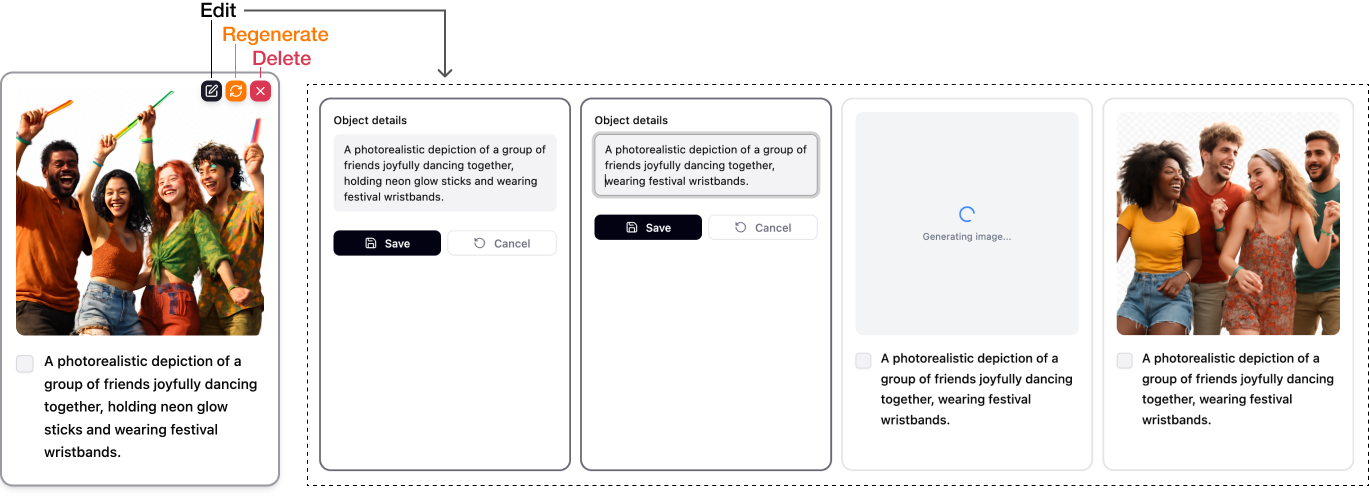}
  \caption{Element card interface in Step 2. The left shows an element card with Edit, Delete, and Regenerate buttons. The right shows the editing interaction flow where users can modify the rough prompt (removing ``holding neon glow sticks'' in this example) and update the preview image.}
  \label{fig:screen_step2_card_ops}
\end{figure}

The user interface (UI) for this step is designed to facilitate smooth interaction with the recommended elements. Users can initiate the process to trigger the AI to generate candidates for all element categories simultaneously. The UI then displays \emph{element cards} corresponding to each category. The cards for \emph{Object}, \emph{Background}, \emph{Typography}, and \emph{Composition} include the preview images generated by ElementVisualizer along with the rough prompts, while the \emph{Text} card presents textual information in the format ``Role Name: Text.'' At the category level, users can perform \emph{Add} and \emph{Extra Recommendation} operations. The Add operation allows users to manually create new element candidates, while Extra Recommendation reinvokes the ElementRecommender to generate additional candidates. At the individual card level, users can perform \emph{Edit}, \emph{Delete}, and \emph{Regenerate} operations (Figure~\ref{fig:screen_step2_card_ops}). The editing operation is performed at the rough prompt level, and once changes are confirmed, ElementVisualizer automatically restarts to create new enhanced prompts and corresponding preview images. The regenerating operation allows users to restart ElementVisualizer based on the same rough prompt to generate subtle variations. This enables users to smoothly perform iterative adjustments and visual exploration at the element level.

Users can select their preferred candidates from the generated element cards to form a \emph{selected set of elements}, which serves as the basis for the final integrated generation step. The selected elements are visually highlighted within the interface, and users can freely modify or replace them. Through this iterative interaction, users progressively refine and align the visual components with their design intentions, guided by AI-generated suggestions.
Rather than functioning as a simple preparatory stage for image generation, this step serves as an interactive design workspace that supports exploration, recomposition, and refinement at the element level. Although the generation of preview images is not performed in real time, it is designed to be lightweight and fast enough to support continuous user engagement and smooth iteration. The structured card-based interface, combined with this rapid visual feedback, enables users to externalize and evaluate their ideas efficiently. In doing so, users engage in a collaborative creative dialogue with AI, transforming abstract design intents into concrete, visually grounded concepts.

\subsection{Step 3: Final Design Generation via Element Integration}

\begin{figure}[t]
  \centering
  \includegraphics[width=\columnwidth]{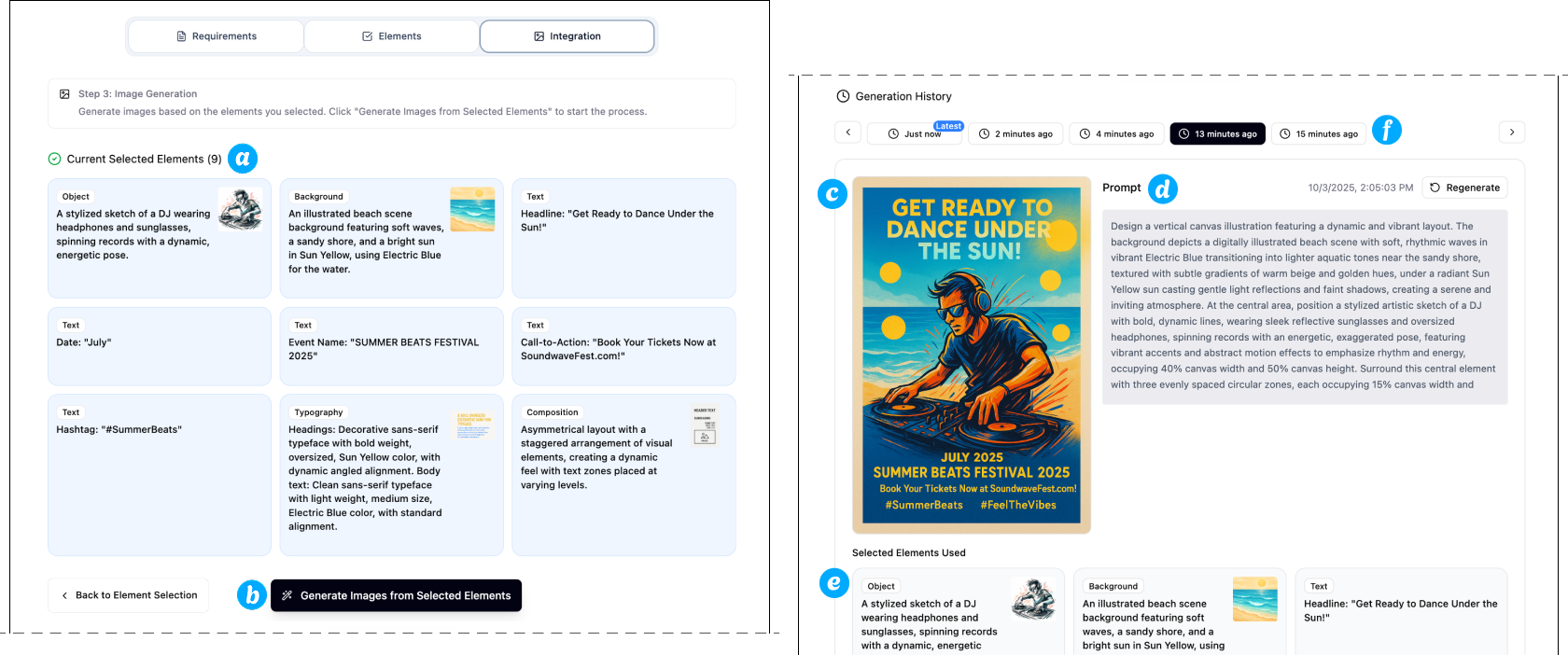}
  \caption{Step 3 interface: Integrating selected elements into a final design. Users review their selected elements~(a) from Step 2 and initiate integration generation~(b). The system then displays the generated final design~(c), the integrated prompt~(d), and the used elements~(e). Users can explore variations through the generation history~(f).}
  \label{fig:screen_step3}
\end{figure}

This step focuses on the process where the AI integrates the user-selected design elements and generates the final design image (Figure~\ref{fig:screen_step3}). The goal is to combine the individual elements (Objects, Backgrounds, Text, Typography, Composition) formed in Step 2 into a coherent design output that maintains the relationships between elements within the overall structure. This step involves AI referencing the previously defined and adjusted element information comprehensively while considering layout, hierarchy, and text placement to generate a high-quality visual.

The core of this step is the \emph{DesignIntegrator} module, which takes the enhanced prompts of the user-selected elements as input and generates a final \emph{integrated prompt} for image generation. DesignIntegrator employs a composition-first approach, using the Composition element as the structural foundation and sequentially incorporating other elements. The process follows a four-stage integration procedure defined in the system prompt:
\begin{enumerate}
    \item \textbf{Background Integration:} Retain the Composition as the layout skeleton and apply the characteristics of the Background element (patterns, color tones, textures, etc.) within it. The division and geometric arrangement of the background structure are preserved based on the Composition specifications.
    \item \textbf{Text Integration:} Insert the content of the Text element into the text areas while considering their roles, \eg, heading. The numerical specifications of the text areas, such as size and coordinates (\eg, ``top 25\%'', ``width 60\%''), are strictly maintained, but the areas can be proportionally expanded as needed to prioritize readability.
    \item \textbf{Typography Enhancement:} Integrate the characteristics of the Typography element (font style, weight, color, etc.) into the integrated text naturally. Any other visual content is discarded to avoid disrupting the spatial structure of the Composition while enhancing visual unity.
    \item \textbf{Object Placement:} Insert the content of the Object element into the object areas defined in the Composition (\eg, ``40\% of canvas width'', ``centered''). The numerical specifications of the Composition are prioritized for placement, while the number and scale of elements are adjusted within spatial constraints.
\end{enumerate}
The DesignIntegrator consistently applies three main principles during these processes: ``Composition-First Foundation,'' ``Spatial Precision Maintenance,'' and ``Text Readability Priority.'' The final integrated prompt is then used as input for image generation models.

The user interface for this step is designed to facilitate smooth interaction with the integrated generation process. Users first review the selected elements from Step 2, which are displayed at the top of the screen. They can modify or replace any elements in the previous step if needed. Once satisfied, users can initiate the integrated generation by clicking the ``Generate'' button. The system then processes the input and generates the final design image, taking 60 to 90 seconds. After generation, users can view the resulting design, the integrated prompt used, and the corresponding elements used in the process. Users can also perform a ``Regenerate'' operation to explore subtle variations based on the same set of elements, allowing them to fine-tune for their preferences. Users can also review the generation history, comparing past results to evaluate different design directions.

This step serves as the culmination of the collaborative design process, where the structured elements and user intentions are synthesized into a coherent visual output. The integration generation is designed to be robust and reliable, ensuring that the final design reflects the user's choices while maintaining visual harmony and clarity. The generated design can be used as a final deliverable or as a basis for further refinement and iteration by revisiting previous steps. In summary, this step serves as a robust process for generating designs, translating users' structured considerations into a polished visual outcome.

\subsection{Implementation Details}

\systemname{} is implemented as a web-based application. The frontend is built using React, providing an interactive UI for operations and card management. The backend is based on Python (FastAPI), orchestrating the various AI modules for requirement analysis, element recommendation, and integration generation. Log data, such as generated prompts and edit histories of element cards, is stored in Cloud SQL (PostgreSQL), while generated images are uploaded to Cloud Storage. We use LangChain to interface with OpenAI's GPT-4o for LLM tasks and OpenAI's gpt-image-1 for image generation. The prompts generated at each step, from requirement structuring to final generation, are used as inputs for the respective API calls. The complete system prompts for all modules are provided in Appendix~\ref{sec:appendix-prompts}. This setup allows for a seamless pipeline that integrates LLM-based prompt generation with image generation APIs.

\section{User Evaluation}
We conducted a within-subjects experiment to compare our proposed system with a baseline system that simulates a common conversational AI interface, similar to ChatGPT. The experiment aimed to assess how effectively \systemname{} operationalizes the design goals
relative to the baseline interaction style.

We carefully designed the tasks to simulate a realistic design scenario where participants needed to create multiple design proposals for clients. We collected both system logs and survey responses to evaluate the effectiveness of our proposed system in supporting the design process. 

\subsection{Participants}
We recruited 12 designers (P1-12; 7 women and 5 men, age: M = 27.1, SD = 6.8) from our company. Their design experience ranged from 5 months to 20 years (M = 4.2 years, SD = 5.6). They mainly worked on graphic design tasks such as banners and digital advertisements (12 out of 12), posters, flyers, and direct mail (2), catalogs, brochures, and booklets (2), websites (2), slide materials and presentations (1), and video-related tasks (1). Regarding their experience with GenAI tools, such as ChatGPT and image generation AI, six participants reported using them daily, four reported occasional use, and two reported having used them but not frequently. The designers participated in the study as part of their regular work and were not compensated in addition to their normal pay.

\subsection{Baseline System}

\begin{figure}[t]
  \centering
  \includegraphics[width=0.45\columnwidth]{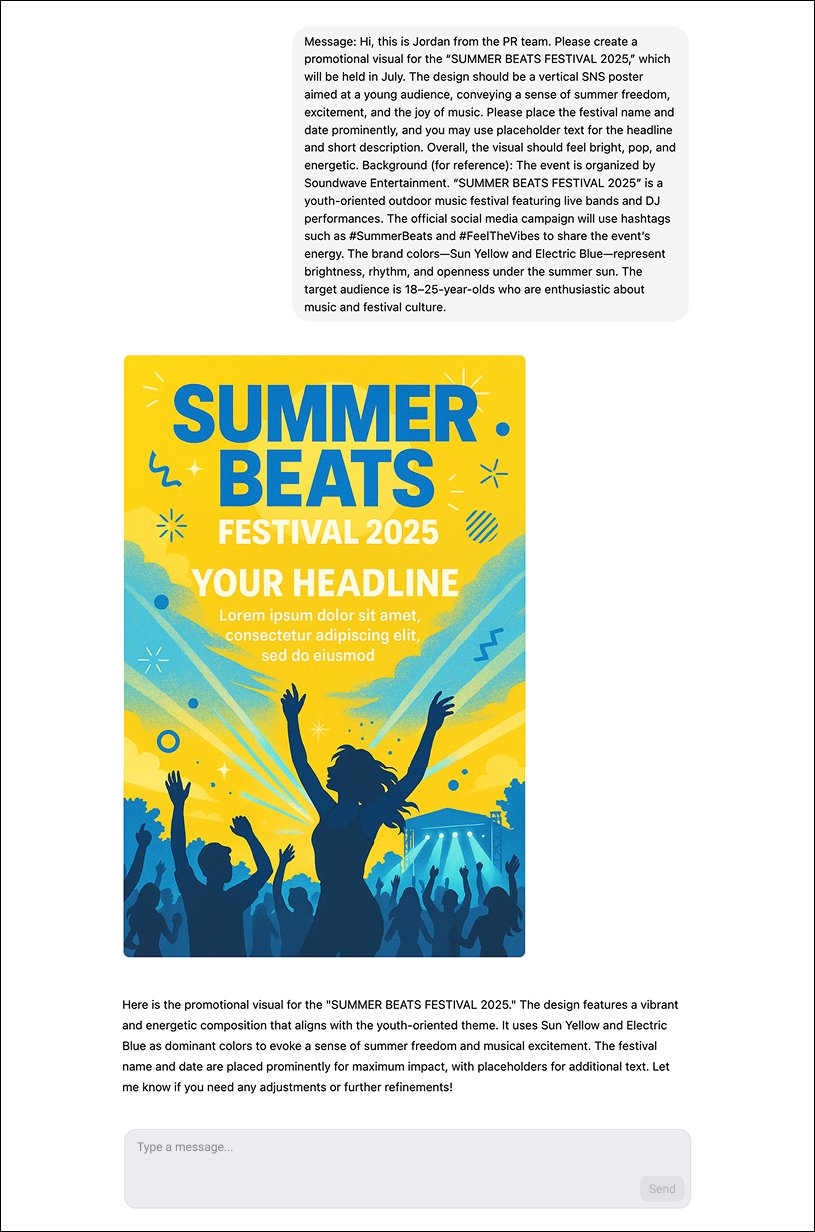}
  \caption{Baseline system interface: A conversational AI interface similar to ChatGPT. Users interact with the AI through text prompts to generate images. The interface features a chat-based interaction model with message history and image generation capabilities.}
  \label{fig:screen_baseline}
\end{figure}

We employ a baseline system that simulates a common conversational AI interface, similar to ChatGPT, which is widely used for design tasks (Figure~\ref{fig:screen_baseline}). The user can interact with the AI model through text prompts to brainstorm ideas, generate images, and refine designs. For the more specialized tools, such as Canva, we decided not to use them as a baseline because they offer a wide range of features, including text editing capabilities and numerous templates, making it challenging to conduct a fair comparison with our proposed system.
We implemented the ReAct agent~\cite{YaoReAct2023} using LangGraph~\cite{LangChainLangGraph2024}, which has the image generation model as a tool. We implemented a chat interface based on agent-chat-ui~\cite{LangChainAgent2025}, an open-source implementation for LangGraph agents. We used the same image generation model and LLM as in our proposed system. Since the baseline system does not support generating multiple images simultaneously, we imposed a similar restriction on our proposed system, \ie, generating design images one by one, for a fair comparison.

\subsection{Tasks and Design}
We designed a within-subjects experiment where participants used both the baseline system and our proposed system to complete design tasks. The design tasks simulated a scenario where participants received a design request from a client and needed to present multiple design proposals to the client, each with a different direction or style to discuss and align on the design direction with the client.
Participants were instructed to create a total of two images using the displayed tool, following the specified design conditions. The participants were asked to satisfy the client's requirements and specifications and design freely for the parts not specified by the client. We set a 15-minute time limit for each task to ensure a reasonable experiment duration. For the image submission, participants were instructed to submit two images when they felt they had created ``two sufficiently different design proposals that meet the client's requirements and specifications'' or when there was only one minute remaining in the task. 

We created two types of task conditions with varying levels of design freedom. One was a ``low-freedom task condition,'' and the other was a ``high-freedom task condition,'' with two tasks each, for a total of four tasks (the full task briefs are provided in Appendix~\ref{sec:appendix-tasks-and-outputs}).
Each participant was assigned to complete two tasks under each condition, using two different systems. The order of the four tasks was fixed, while the order of the systems was counterbalanced within each task condition.

\subsection{Procedure}
We conducted all the experiments remotely. We first conducted an initial briefing session via video call, where we explained the experiment procedure and the functions of each tool. After that, participants signed a consent form. Next, we provided them with access to practice each tool and asked them to familiarize themselves with the tools using practice tasks. After they felt sufficiently comfortable with the tools, we asked them to report the time they spent practicing (Baseline: M = 18.6 min, SD = 9.2; Proposed: M = 22.6 min, SD = 12.7). We also asked them to complete a demographic survey at this time.

Once we confirmed that they had submitted the practice time report and demographic survey, we provided them access to the main design tasks. For each task, they accessed the experiment system using a provided URL, read the detailed task instructions, and then used the displayed tool to create images according to the task requirements within the time limit. The experiment system has a countdown timer and a drag-and-drop form for submitting images. When the time limit was reached, the system automatically stopped the task and submitted the images registered in the form. After submitting their images, they completed a post-task survey about their experience using the tool for each task (explained later). They are prompted to complete the tasks in the specified order and to work on each task without breaks until they complete the post-task survey. They are also instructed not to use external tools such as Google search or ChatGPT and not to reference any materials other than those provided in the task instructions. Due to the limited text drawing capabilities of the image generation, we instructed them to submit their images as they were, even if the rendered text contains distortions or errors, as we considered that it would not affect the alignment of the design direction with the client. The experiment, including the practice session, took approximately 120 minutes to complete.

The post-task survey includes three types of questions.
The first type is the Creativity Support Index (CSI)~\cite{Cherry2014-vs}, which measures the extent to which a tool supports creativity. We employ a simplified version of the CSI, excluding the Collaboration factor and the weighting questions, resulting in a total of 10 questions across five factors: Enjoyment, Exploration, Expressiveness, Immersion, and Results Worth Effort. Each question is rated on a Likert scale from 0 (Strongly Disagree) to 10 (Strongly Agree), with 5 representing a neutral stance.
The second type is questions specific to the proposed system, asking participants to rate the usefulness of each function in the proposed system. There are nine questions in total, rated on a Likert scale from 1 (Not at all useful) to 5 (Very useful), with a neutral stance represented by 3.
The last type is an optional, open-ended question that asks participants to share their thoughts and experiences using the tool.

\subsection{Data Collection and Analysis}
We collected two types of data: system logs and survey responses. The system logs include the time of each image generation and submission, the generated images and their corresponding prompts, the submitted images, and the task completion status. We define the task completion status as whether the participant submitted two images, regardless of whether it was within the time limit or automatically submitted when the time ran out. Note that the element-level preview images generated in the proposed system were not logged, as they were intended solely for reference and not for submission. The survey responses include Likert-scale ratings for CSI (ranging from 0 to 10) and for the functionality of the proposed system (ranging from 1 to 5), as well as optional answers to an open-ended question about their experience using the system.

We compared the numerical differences between the task completion rate and time, as well as the number of images generated. We also compute diversity scores for a set of generated images and prompts. The diversity score is defined as the average pairwise dissimilarity between all pairs of items in a set. We compute the dissimilarity with $1 - \textrm{cos}(\mathbf{x}_i, \mathbf{x}_j)$, where $\mathbf{x}_i$ and $\mathbf{x}_j$ are the feature embeddings of items $i$ and $j$, respectively, and $\textrm{cos}(\cdot, \cdot)$ is the cosine similarity function. We use the DreamSim model~\cite{FuDreamSim2023} to extract image embeddings and the OpenAI text-embedding-3-large model~\cite{OpenAIOpenAI2024} to extract prompt embeddings.

For survey data, we have two ratings for each system under each task condition. For the CSI ratings, we averaged them for each system and compared them using a paired t-test. For the functionality ratings of the proposed system, we treat them independently and report the proportion of the ratings for each question.

\section{Results}
\label{sec:results}
In this section, we present the results of our user evaluation, focusing on both quantitative metrics and qualitative feedback from participants. We analyze the performance of our proposed system compared to the baseline system across various dimensions, including task completion rates, time taken, image and prompt diversity, perceived creativity support, and feature usefulness.

\setlength{\tabcolsep}{2pt}
\begin{figure*}[t]
  \centering
  \begin{tabular}{ccccc}
    \multicolumn{2}{c}{Low-freedom tasks} & & \multicolumn{2}{c}{High-freedom tasks} \\[3pt]
    \frame{\includegraphics[width=51pt]{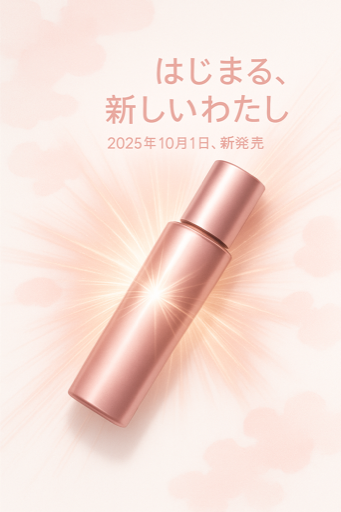}}\frame{\includegraphics[width=51pt]{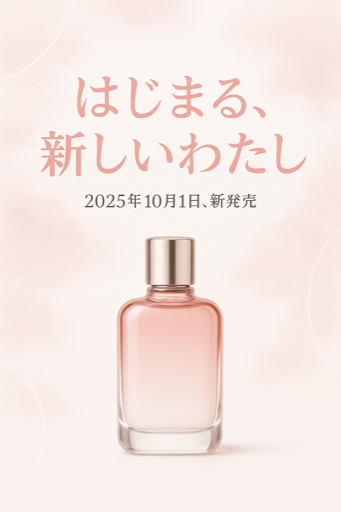}}&\frame{\includegraphics[width=51pt]{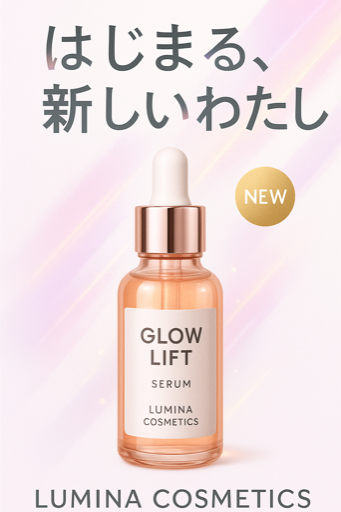}}\frame{\includegraphics[width=51pt]{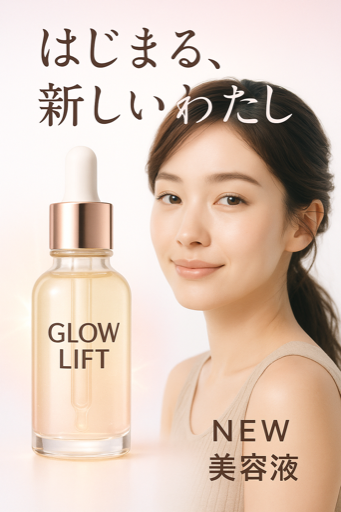}}& &\frame{\includegraphics[width=51pt]{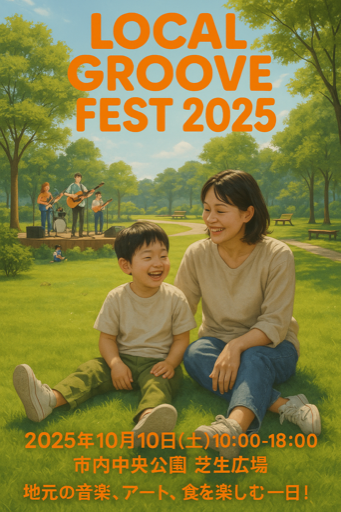}}\frame{\includegraphics[width=51pt]{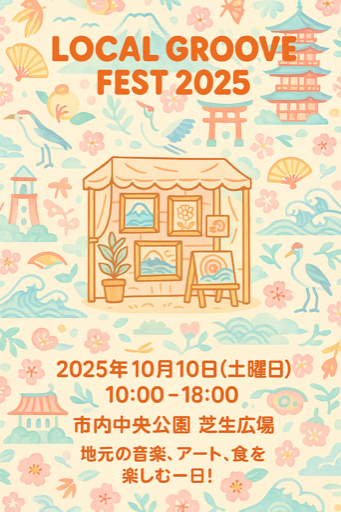}}&\frame{\includegraphics[width=51pt]{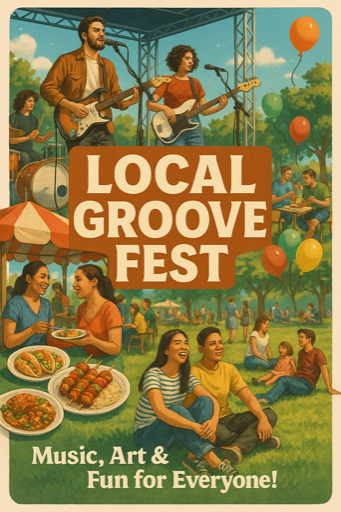}}\frame{\includegraphics[width=51pt]{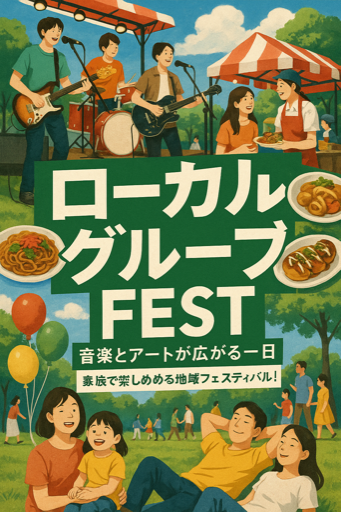}} \\
    P2 with Brief2Design & P11 with Baseline & & P7 with Brief2Design & P6 with Baseline \\
  \end{tabular}
  \caption{Examples of submitted images from different participants (P2, P11, P7, and P6) under different task conditions (low-freedom and high-freedom) and systems (Brief2Design and Baseline).}
  \label{fig:example-outputs}
\end{figure*}
\setlength{\tabcolsep}{6pt}

\subsection{Completion Rate and Time}
The completion rates for the low-freedom tasks were 66.7\% (8/12) with the proposed system and 91.7\% (11/12) with the baseline, while for the high-freedom tasks they were 75.0\% (9/12) and 91.7\% (11/12), respectively.
Among the completed tasks, the average completion time for the low-freedom tasks was 13.8 minutes ($SD = 1.1$) with the proposed system and 11.7 minutes ($SD = 3.5$) with the baseline. In contrast, for the high-freedom tasks, they were 10.9 minutes ($SD = 2.0$) and 11.2 minutes ($SD = 2.9$), respectively. A paired t-test revealed no significant difference between the systems for either the low-freedom ($t(6) = 1.52$, $p = .181$) or high-freedom ($t(7) = -0.31$, $p = .768$) tasks. Note that P4 in the low-freedom task with the baseline system submitted images with an experimenter's help after task completion due to a misunderstanding of the experiment system. Therefore, we omitted this data point from the completion time analysis. Some participants also reported that the image submission process was confusing: \mention{P9; proposed}{I was unsure how to submit the images, so I ended up submitting zero images.}. Some participants mentioned that the generation time was long, which affected their workflow: \mention{P3; baseline}{Since the generation takes time, the number of iterations is inevitably reduced.}, \mention{P5; proposed}{The generation took longer than I expected, and the waiting time was longer than the working time.}, \mention{P12; proposed}{The generation took a long time, and I had to wait for a considerable period. The final generation took nearly 1.5 minutes. To create something accurate within 15 minutes, I would need to produce several variations, which proved to be a bit challenging.} We show four representative pairs of submitted images in Figure~\ref{fig:example-outputs}. The complete set of outputs from all participants is available in Appendix~\ref{sec:appendix-tasks-and-outputs}.

\subsection{Number of Generated Images and Time per Generation}
\label{sec:results_num_images}
We analyzed the number of images generated by participants during the tasks. In the low-freedom tasks, participants generated an average of 4.2 images ($SD = 1.4$) with the proposed system and 4.8 ($SD = 2.1$) with the baseline. In contrast, in the high-freedom tasks, they generated 3.8 ($SD = 0.9$) and 5.4 ($SD = 1.2$), respectively. A paired t-test showed no significant difference in the number of generated images for the low-freedom tasks ($t(11) = -1.07$, $p = .309$), but a significant difference for the high-freedom tasks ($t(11) = -4.71$, $p = .001$).
The average time per image generation, \ie, the completed time divided by the number of generated images, was 3.4 minutes ($SD = 0.8$) and 2.8 ($SD = 0.4$) for the low-freedom tasks, and 3.1 minutes ($SD = 0.7$) and 2.2 ($SD = 0.3$) for the high-freedom tasks, respectively. The differences were marginally significant for the low-freedom tasks ($t(6) = 2.04$, $p = .088$) but were significant for the high-freedom tasks ($t(7) = 3.10$, $p = .017$).

\begin{figure}[t]
  \centering
  \includegraphics[width=\columnwidth]{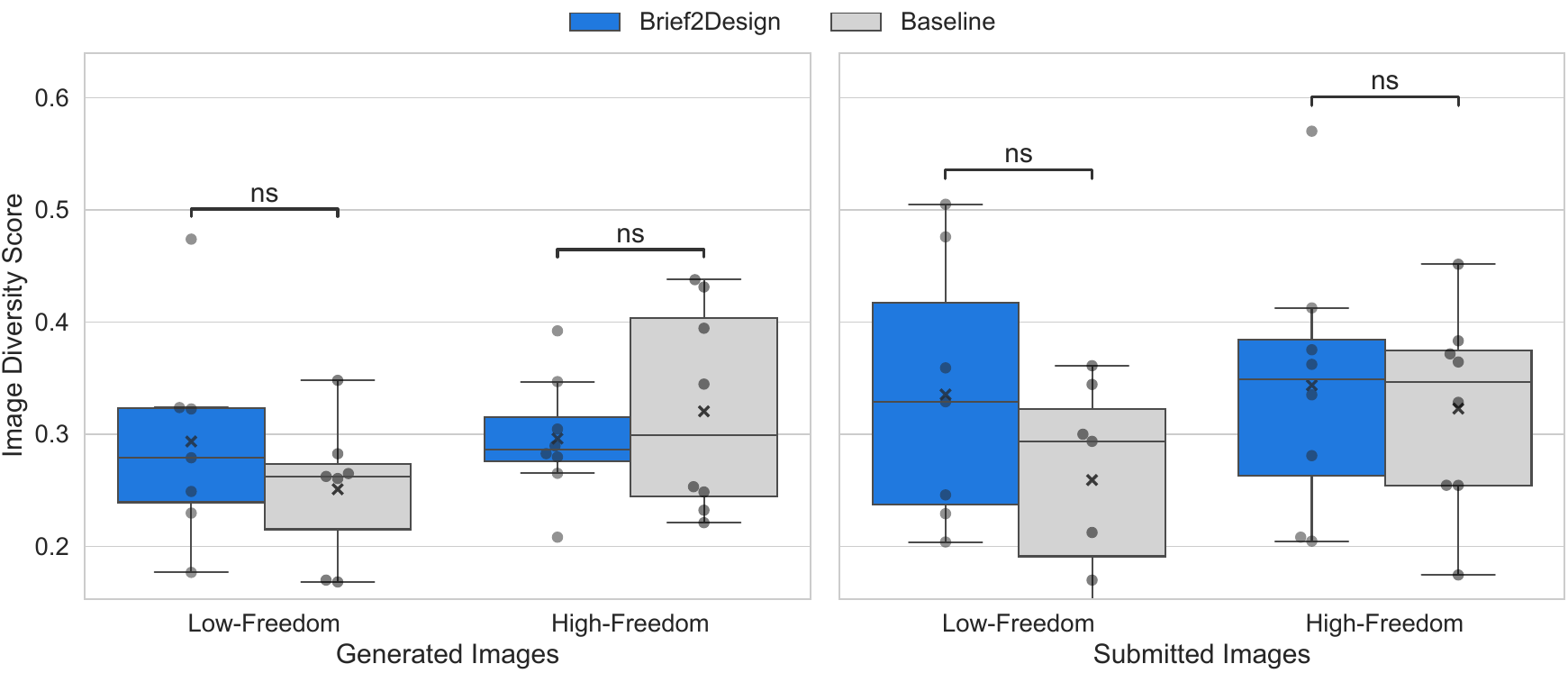}
  \caption{Image diversity comparison for generated and submitted images (ns: not significant). The circle markers indicate data points, and the cross markers indicate means.}
  \label{fig:image_diversity_comparison}
\end{figure}

\begin{figure}[t]
  \centering
  \includegraphics[width=\columnwidth]{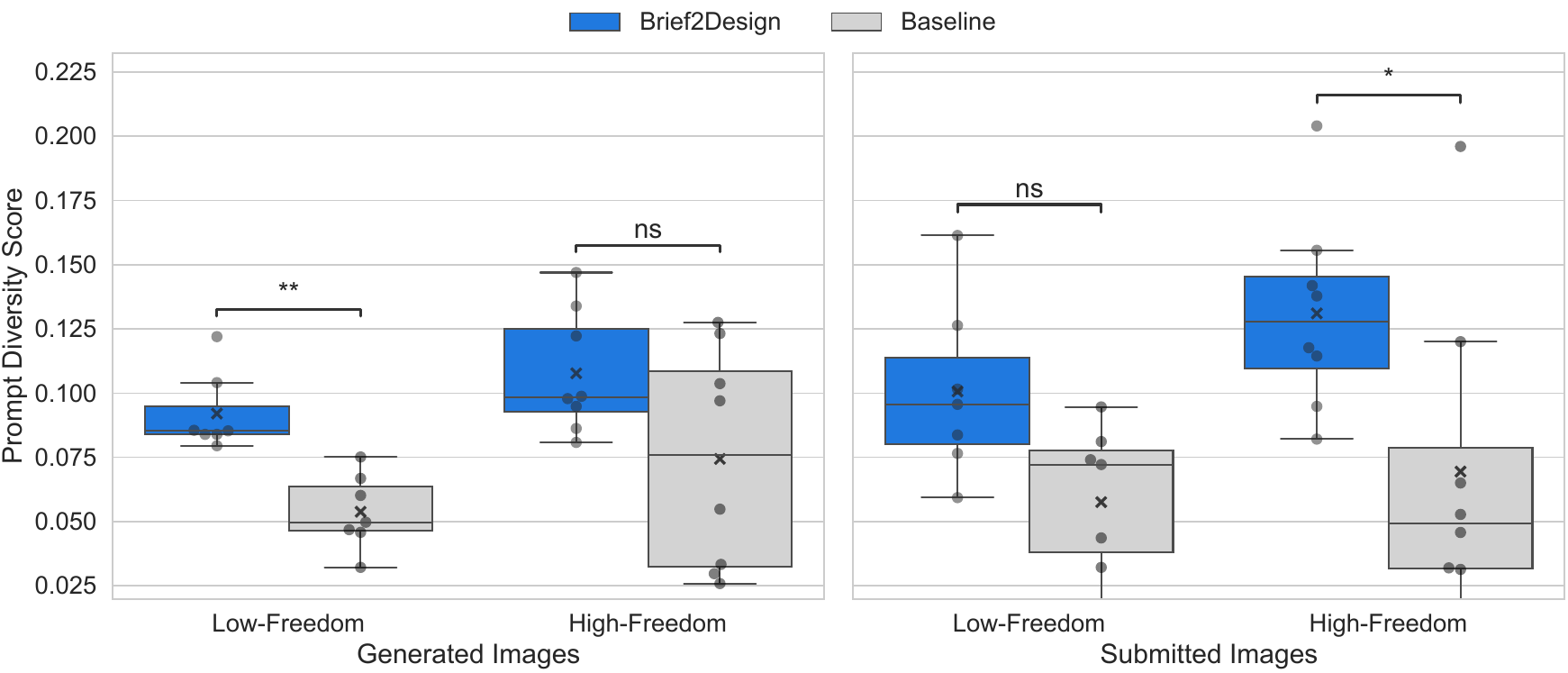}
  \caption{Prompt diversity comparison for generated and submitted images (ns: not significant, $\ast$: $p < .05$, $\ast\ast$: $p < .01$). The circle markers indicate data points, and the cross markers indicate means.}
  \label{fig:prompt_diversity_comparison}
\end{figure}

\subsection{Image and Prompt Diversity}
\label{sec:results_diversity}
\begin{figure}[t]
  \centering

  \begin{minipage}[b]{0.64\linewidth}
    \centering
    \includegraphics[width=\linewidth]{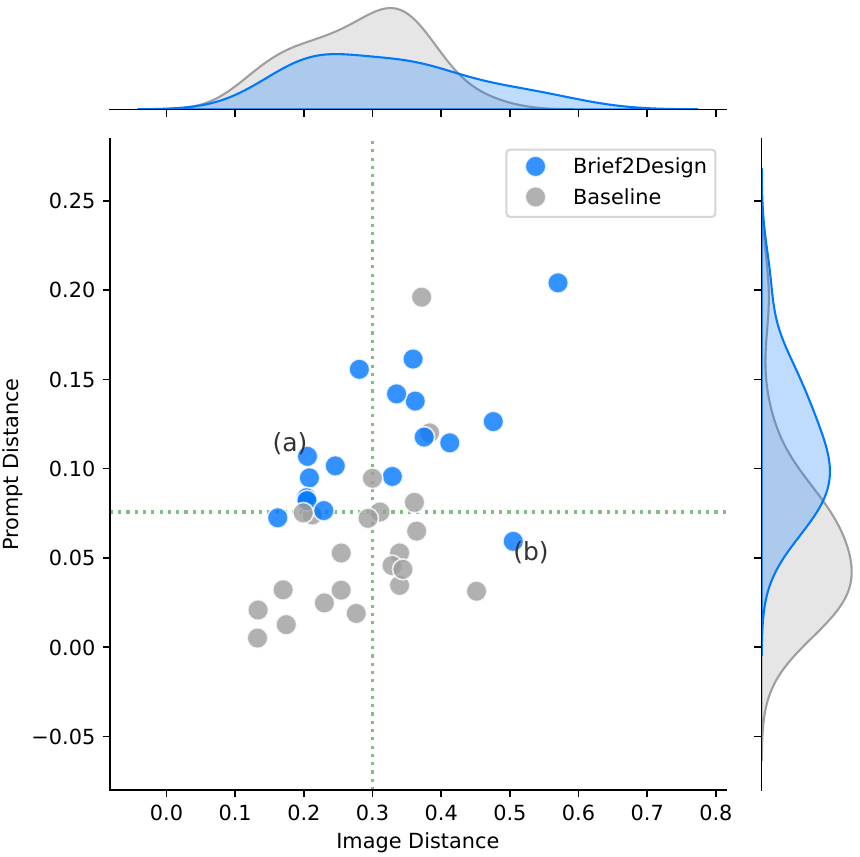}
  \end{minipage}\hfill
  \begin{minipage}[b]{0.32\linewidth}
    \begin{subfigure}{\linewidth}
      \centering
      \frame{\includegraphics[width=.49\linewidth]{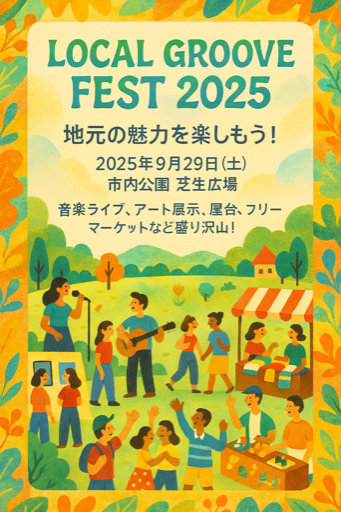}}%
      \hfill
      \frame{\includegraphics[width=.49\linewidth]{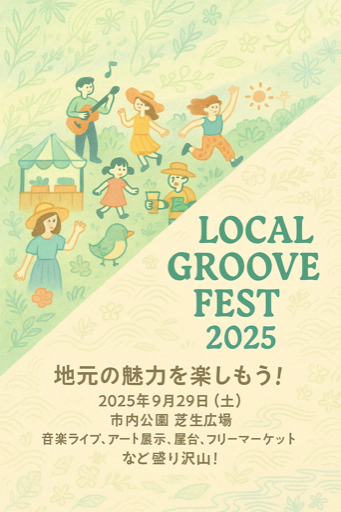}}
      \caption{Higher prompt distance (.107) but lower image distance (.205).}
      \label{fig:pair1}
    \end{subfigure}

    \vspace{0.012\textheight} %

    \begin{subfigure}{\linewidth}
      \centering
      \frame{\includegraphics[width=.49\linewidth]{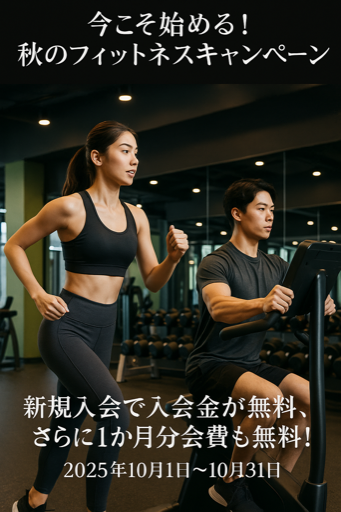}}%
      \hfill
      \frame{\includegraphics[width=.49\linewidth]{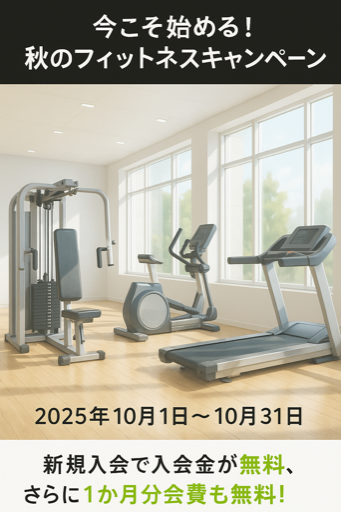}}
      \caption{Lower prompt distance (.059) but higher image distance (.505).}
      \label{fig:pair2}
    \end{subfigure}
  \end{minipage}

  \caption{Left: Scatter plot showing the relationship between prompt distance and image distance for the submitted images. The dotted green lines indicate the medians of each axis. Right: Two example image pairs corresponding to representative points (a) and (b).}
  \label{fig:scatter_with_pairs}
\end{figure}

We computed the image and prompt diversity scores for the sets of images and prompts generated and submitted by participants. For this analysis, we use the paired data of the tasks where participants submitted two images, resulting in 7 data points for the low-freedom tasks and 8 for the high-freedom tasks. The results are shown in Figures~\ref{fig:image_diversity_comparison} and \ref{fig:prompt_diversity_comparison}.
The t-test revealed no significant differences in image diversity across systems for the low-freedom tasks ($t(6) = .824$, $p = .441$) and the high-freedom tasks ($t(7) = -1.16$, $p = .283$) in the generated images, and for the low-freedom tasks ($t(6) = 1.91$, $p = .105$) and the high-freedom tasks ($t(7) = .820$, $p = .439$) in the submitted images.
In contrast, for prompt diversity, there was a significant difference between systems for the low-freedom tasks ($t(6) = 3.86$, $p = .008$) and no significant difference for the high-freedom tasks ($t(7) = 1.69$, $p = .134$) in the generated images, and a marginally significant difference for the low-freedom tasks ($t(6) = 2.00$, $p = .093$) and a significant difference for the high-freedom tasks ($t(7) = 2.45$, $p = .044$) in the submitted images.
We plot the relationship between the prompt distance and image distance for the submitted images in Figure~\ref{fig:scatter_with_pairs} for further analysis.

\subsection{Questionnaire Results}
\label{sec:results_questionnaire}
\begin{figure}[t]
  \centering
  \includegraphics[width=.7\columnwidth]{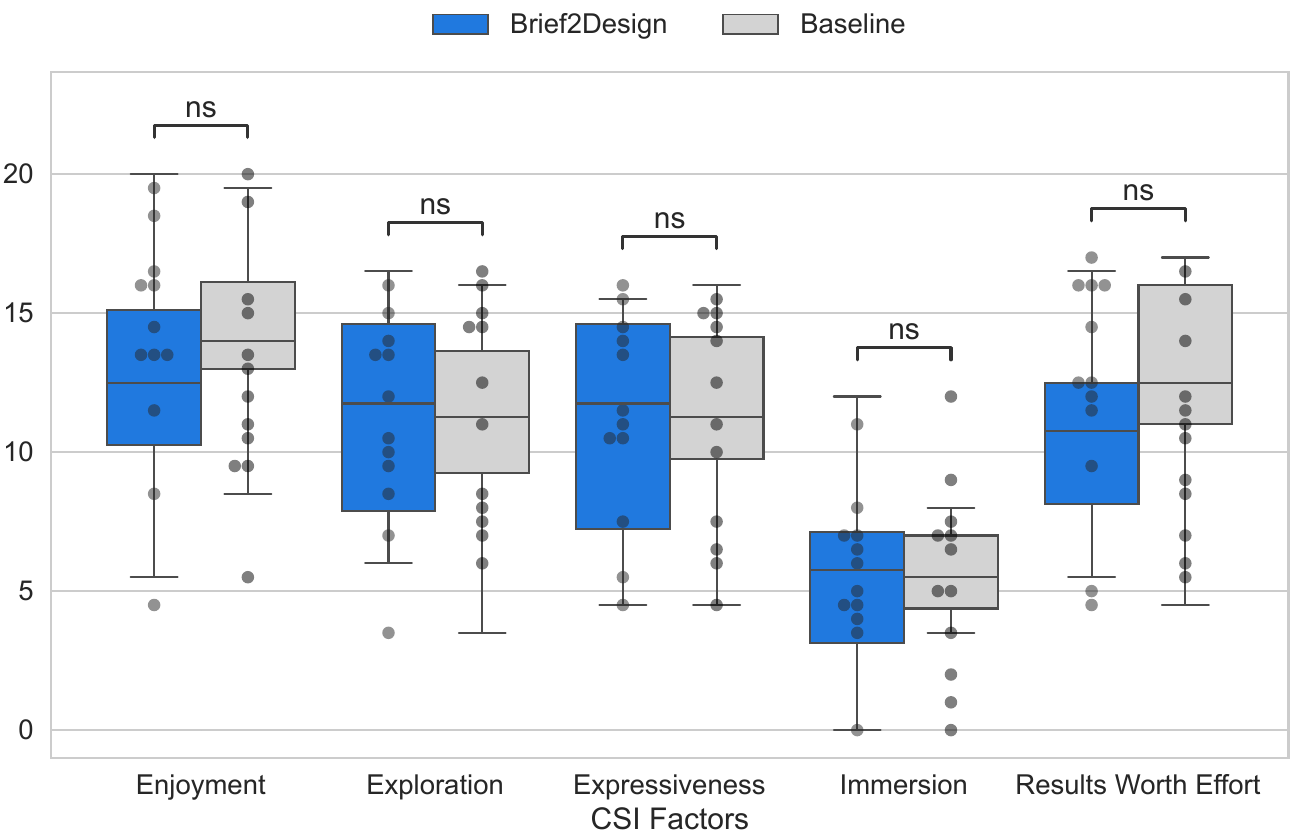}
  \caption{Results of the Creativity Support Index (CSI) questionnaire.}
  \label{fig:questionnaire_csi}
\end{figure}

\begin{figure}[t]
  \centering
  \includegraphics[width=.9\columnwidth]{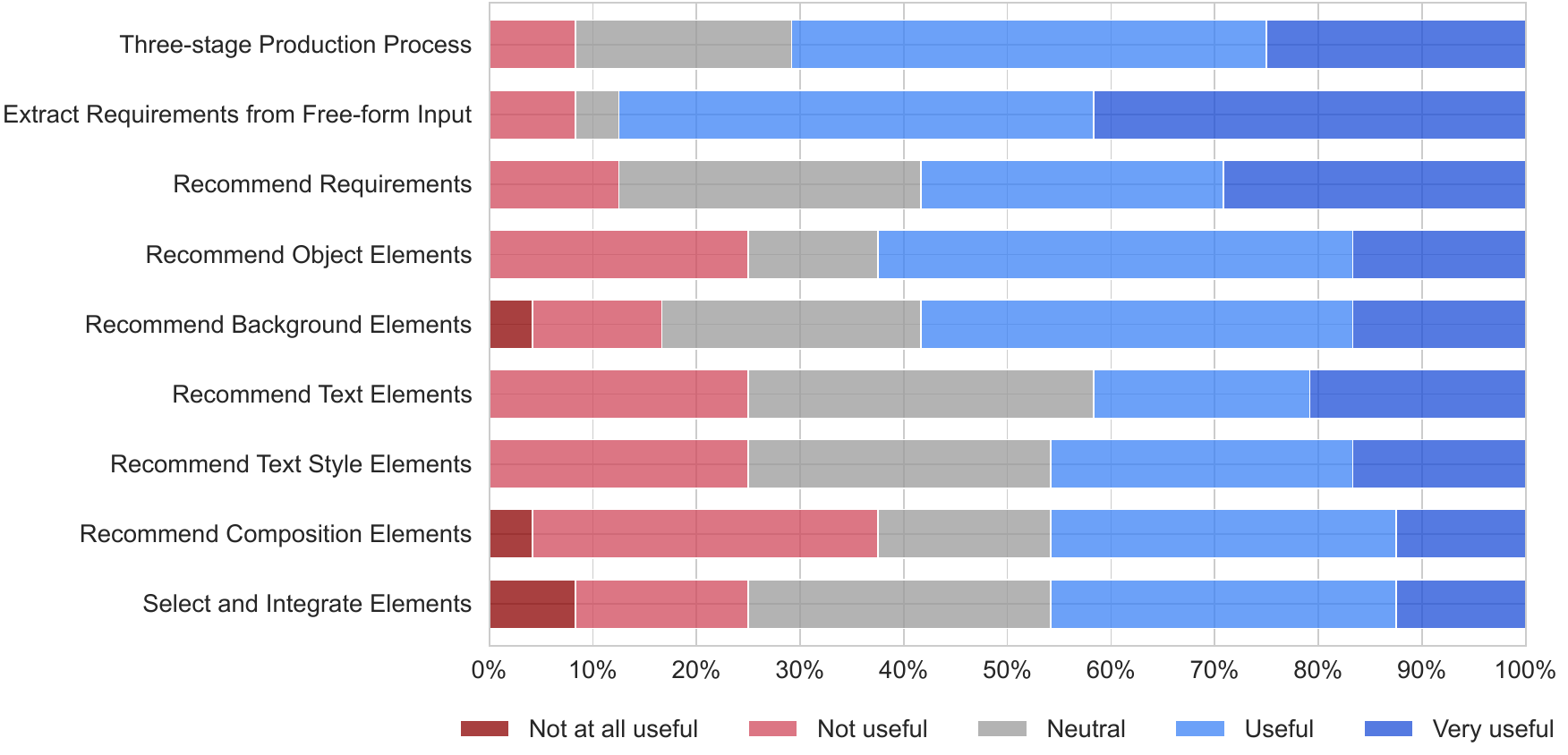}
  \caption{Rating proportions for each feature in the proposed system.}
  \label{fig:questionnaire_feature}
\end{figure}

We analyzed the results of the Creativity Support Index (CSI) questionnaire, as shown in Figure~\ref{fig:questionnaire_csi}. Overall, CSI scores were 51.3 ($SD = 15.3$) for the proposed system and 54.0 ($SD = 15.1$) for the baseline system. The Wilcoxon signed-rank test revealed no significant difference between the systems ($W = 31.0$, $p = .569$). It also showed no significant differences between the systems in any of the five individual factors, Enjoyment ($W = 19.5$, $p = .259$), Exploration ($W = 31.0$, $p = .896$), Expressiveness ($W = 31.0$, $p = .905$), Immersion ($W = 26.0$, $p = .896$), and Results Worth Effort ($W = 21.0$, $p = .167$).
We also evaluated the usefulness of each feature in the proposed system, as shown in Figure~\ref{fig:questionnaire_feature}. Participants rated the features preferably as follows: \textit{Three-stage Production Process} ($M = 3.9$, $SD = .90$), \textit{Extract Requirements from Free-form Input} ($M = 4.2$, $SD = .88$), \textit{Recommend Requirements} ($M = 3.8$, $SD = 1.03$), \textit{Recommend Object Elements} ($M = 3.5$, $SD = 1.06$), \textit{Recommend Background Elements} ($M = 3.5$, $SD = 1.06$), \textit{Recommend Text Elements} ($M = 3.4$, $SD = 1.10$), \textit{Recommend Text Style Elements} ($M = 3.4$, $SD = 1.06$), \textit{Recommend Composition Elements} ($M = 3.2$, $SD = 1.17$), and \textit{Select and Integrate Elements} ($M = 3.2$, $SD = 1.15$).

\section{Discussion}

\subsection{Brief2Design Takes Pre-Generation Time but Yields Diverse Outputs.}
\label{sec:discussion_time_diversity}
Prior research has shown that structured, pre-generation scaffolding can enhance output diversity and quality despite requiring additional time~\cite{Tao2025-zu, Feng2024-em}.
DesignWeaver~\cite{Tao2025-zu} found that participants who spent more time structuring prompts through dimensional tagging produced semantically more diverse images, while PromptMagician~\cite{Feng2024-em} demonstrated that pre-generation exploration of style modifiers led to higher aesthetic satisfaction.

We observed similar trade-offs in our study.
Participants took more time per generation with Brief2Design compared to the baseline system in both task conditions (see Section~\ref{sec:results_num_images}), likely due to organizing requirements and previewing element-level images before generating final designs.
However, participants with Brief2Design generated fewer images in the high-freedom condition, suggesting that the system enabled users to achieve sufficient diversity with fewer generation attempts when requirements were less constrained.
Brief2Design's structured approach allows quick recombination of design elements across multiple categories (objects, backgrounds, text, typography, composition), potentially reducing trial-and-error iterations needed to achieve diversity.

Note that the lack of generation count difference in the low-freedom condition may reflect the diversity of Baseline's outputs plateauing early with constrained requirements, causing participants to finish earlier without attempting additional generations.

\subsection{Compositional Diversity and Evaluation Metrics}
\label{sec:discussion_compositional_diversity}
As discussed in Section~\ref{sec:results_diversity}, both the generated and submitted images did not show significant differences in image diversity across systems. However, prompt diversity showed significant differences in the low-freedom condition for generated images ($p = .008$) and marginally significant or significant differences in submitted images across both conditions.

This divergence between prompt diversity and image diversity suggests that perceptual similarity metrics, while effective for capturing semantic and visual differences, may not adequately capture compositional variations such as layout and spatial arrangement that are central to graphic design. Figure~\ref{fig:scatter_with_pairs} illustrates cases where higher prompt distance does not correspond to higher image distance, particularly when variations involve compositional rather than semantic changes.

Participants' qualitative feedback suggests that compositional exploration was valued. For example, \mention{P7}{Specifying the composition not in words but intuitively is really good!}. However, as discussed in Section~\ref{sec:results_questionnaire}, some participants found compositional recommendations challenging to evaluate, suggesting areas for improvement in preview fidelity and interface design.

These observations highlight the need for evaluation metrics that account for compositional diversity in graphic design contexts, where layout and structure play a central role alongside visual appearance. Future work could explore layout-specific similarity measures~\cite{OtaniLTSim2024} for more appropriate evaluation of compositional variations.

\subsection{User Perceptions and Underlying Factors}
\label{sec:discussion_user_perceptions}

We did not find significant differences in overall creativity support between the two systems as reported in Section~\ref{sec:results_questionnaire}. However, participants rated Brief2Design's features positively, with the three-stage production process ($M = 3.9$) and requirement extraction ($M = 4.2$) receiving particularly high ratings. Among the recommendation features, Requirements ($M = 3.8$), Object ($M = 3.5$), and Background ($M = 3.5$) were well-received, while Text ($M = 3.4$), Text Style ($M = 3.4$), and Composition ($M = 3.2$) received relatively lower ratings. The Select and Integrate Elements feature ($M = 3.2$) also received relatively lower ratings.

This pattern may reflect the effectiveness of these features in supporting early design exploration. Requirements, objects, and backgrounds help users define and explore foundational aspects of their designs. Requirements clarify design intent through textual suggestions, while visual elements (objects, backgrounds) provide concrete starting points that participants could evaluate through preview images. These features support conceptual exploration during early design stages, with previews offering tangible reference points that make recommendations easier to assess and select.

In contrast, text content, typography, and composition involve spatial and stylistic decisions that are more difficult to evaluate in isolation. Some of these elements are inherently context-dependent: as P2 noted, text style choices depend on ``various other elements (background, objects, overall atmosphere),'' making isolated preview images less helpful for decision-making. Similarly, participants raised concerns about preview quality for abstract design dimensions like composition, suggesting difficulties in evaluating these elements effectively and creating a gap between user expectations and what recommendations could provide.

Beyond individual element evaluation, the integration process posed additional challenges. Participants reported inconsistencies between element previews and final outputs: P8 noted that objects ``with completely different atmosphere from the selection phase'' appeared in final generation, while P7 found that selected fonts were not reflected accurately. P2 raised concerns about composition previews, noting that placeholder objects shown in layout previews might unintentionally influence final generation: ``I was worried that if I chose this composition, the final output would be incredibly biased in this way.'' Some participants also experienced element compatibility issues, with P3 observing that ``each element became separate and the final design felt disjointed overall,'' and P7 questioning whether ``the elements I chose had poor compatibility.'' The lower rating for the integration feature ($M = 3.2$) reflects these coordination challenges, suggesting that final assembly requires additional interface support to maintain coherence across selected elements.

Despite these challenges with certain features, participants generally responded positively to Brief2Design's approach, as reflected in the feature ratings (Figure~\ref{fig:questionnaire_feature}). The high ratings for the three-stage process and requirement extraction suggest that the structured workflow's core benefits outweighed the difficulties encountered with specific element types. P2 noted that generating multiple elements at once ``provided detailed idea variations that became a great stimulus for creation,'' suggesting that the element-based exploration, while requiring adaptation, can inspire creative exploration and support designers in generating diverse concepts.

\subsection{Workflow Alignment Challenges}
\label{sec:discussion_workflow_alignment}

One participant (P2) provided critical feedback on workflow alignment: \textit{``I felt that the process of creating a design with this tool was quite different from my usual workflow. In my regular design process (without AI), I tend to refine all elements gradually, like stacking blocks in a balanced way. In contrast, the guide-based approach feels more like integrating and adjusting disparate elements all at once at the end. When I actually used this tool, I found that the sensation of creating was quite different, and I struggled a bit to get used to it during practice. However, once I generated an image, the feeling became closer to my usual adjustment phase. With time, familiarity, and some tips on prompt input, I believe the quality will improve.''}

This feedback reveals a tension between Brief2Design's staged workflow and the gradual refinement approaches some designers prefer. While the structured workflow benefits requirement clarification and systematic exploration, it requires users to adapt to a different creative process. This reflects the broader challenge of workflow reconstruction when integrating generative AI tools~\cite{SimkuteIronies2024}, where new interaction patterns may initially feel unfamiliar even when offering long-term benefits. P2's acknowledgment of potential quality improvement with familiarity suggests that structured exploration, while initially challenging, may become valuable as users develop fluency with the tool.

\subsection{Design Implications}

\subsubsection{Front-Load Creative Decisions for Deliberate Exploration}

As observed in Section~\ref{sec:discussion_time_diversity}, Brief2Design's structured workflow required more time per generation but enabled users to achieve diverse outputs with fewer generation attempts in less constrained tasks. This aligns with prior research showing that guiding users to explore conceptual dimensions before generating complete outputs helps avoid premature convergence~\cite{Suh2024-de, Tao2025-zu}. For creativity support tools involving generative AI, this suggests that pre-generation structuring should be viewed as deliberate exploration that reduces trial-and-error iterations, rather than as overhead, particularly in less constrained contexts where the risk of early fixation is higher.

\subsubsection{Evaluate Compositional Variations Beyond Perceptual Metrics}

As noted in Section~\ref{sec:discussion_compositional_diversity}, prompt diversity showed significant differences while image diversity (measured by perceptual similarity using DreamSim) did not, revealing a fundamental limitation in how perceptual metrics capture compositional variations such as layout and spatial arrangement. For systems supporting compositional exploration in graphic design, this suggests the need to employ layout-specific similarity measures~\cite{OtaniLTSim2024} alongside perceptual metrics to appropriately evaluate diversity in both visual appearance and compositional structure.

\subsubsection{Bridge the Component-Holistic Gap}

As discussed in Section~\ref{sec:discussion_user_perceptions}, participants rated recommendation features differently based on their role in the design process. Features supporting early design exploration (requirements, objects, backgrounds) received higher ratings than refinement elements (text, typography, composition), and the integration feature also received relatively lower ratings. This pattern reflects two underlying factors. First, context-dependent elements are difficult to evaluate in isolation. Second, there is a gap between element previews and final integrated outputs. For systems that decompose generation into element-level exploration, bridging this component-holistic gap requires (1) providing context-aware previews that show how elements interact with others, and (2) improving preview fidelity to better match final outputs, particularly for abstract design dimensions like composition and typography.

\subsubsection{Balance Structured Guidance with Flexible Workflows}

As revealed in Section~\ref{sec:discussion_workflow_alignment}, some participants felt Brief2Design's staged workflow differed from their gradual refinement approaches. This tension between structure and flexibility has been observed across AI-assisted creative tools, where collaboration modes shift between system-guided and designer-guided approaches depending on task phase and user expertise~\cite{YouDesignManager2025, Tao2025-zu}. Rather than requiring users to adapt to a fixed workflow, systems should balance structured guidance with flexibility by offering multiple pathways through the design process, such as allowing users to skip or reorder stages based on their expertise, or adapting the level of scaffolding to user proficiency.

\subsection{Limitations}

This study has several limitations related to both the user experiment and the system implementation.

\subsubsection{Experimental Setup}
The user experiment was structured around four fixed design tasks, each performed on a different system, with a 15-minute time limit per task and a submission limit of two outputs.  
Compared to real-world professional design workflows, this setup imposed shorter time spans and lower task flexibility, which may have influenced participants' exploration behaviors.  
Therefore, caution should be taken when generalizing the findings to broader design practices.
All twelve participants were professional designers from the same company, with an average of 4.2 years of experience in the field.  
This ensured practical validity but may also introduce bias due to shared organizational practices and design culture.  
On the other hand, the formative study (D1-D6) and the main user experiment (P1-P12) involved separate participant groups, which helped prevent overlap of findings between the exploratory and evaluative phases.

\subsubsection{System Constraints}
In the experiment, both systems were restricted to generate one image at a time to ensure fair comparison with the baseline.
This restriction was introduced for experimental control, and in practical use, multiple or parallel generations could be supported.
The system utilized a high-quality image generation model, resulting in an inference time of approximately 90 seconds.
While this configuration prioritized output quality, faster inference could be preferred in certain use cases, and ongoing improvements in generation technology are expected to mitigate this limitation.

\subsubsection{Interpretation of Findings}
The findings of this study are based on controlled experimental conditions.
Specifically, the limited task duration and homogeneous participant group may constrain the external validity of the results.
Future work should examine the system in more diverse and realistic contexts to better understand how structured workflows function across different design scenarios and user populations.

\section{Conclusion}
In this study, we propose \systemname{}, a T2I-based system that supports a three-step process of brief analysis, element exploration, and integrated generation based on design goals extracted from a formative study ($N = 6$). The structured interaction that encourages requirement clarification enabled multifaceted design exploration, which was previously difficult in conventional conversational workflows.
Through a user study ($N = 12$, tasks with different degrees of freedom) comparing it with a chat-based baseline, we found that \systemname{} encouraged broader prompt-level exploration and was valued for clarifying requirements. However, the additional preparation steps led to longer pre-generation times, lower completion rates under strict time limits, and comparable image diversity. The three-step process and requirement extraction features were highly rated, confirming the effectiveness of brief clarification and guided exploration support.
Future work should examine the system in more diverse and realistic contexts, while addressing challenges in faster inference, improved preview fidelity for context-dependent elements, and flexible stage transitions to better align with diverse design workflows.

\section*{Acknowledgments}
We thank the participants of our formative study and user evaluation for their time and valuable insights.
\paragraph{Use of Generative AI}
We used generative AI tools throughout the project as follows. ChatGPT GPT-5 served as an ideation partner across all stages, including problem framing, study design, iteration on the system concept, and authoring the pseudo-briefs used in the user study. The code for the experimental pipeline was largely prototyped and refined using AI-assisted tools, primarily Anthropic Claude Code and ChatGPT Codex, followed by human verification and integration. The frontend mockups were drafted in Figma Make, and Notebook LM supported literature and interview reviews. During paper writing, we relied on GitHub Copilot for in-editor suggestions and on ChatGPT GPT-5/Codex for drafting and editing paragraphs, with manual review to ensure accuracy and consistency.

\bibliographystyle{ACM-Reference-Format}
\bibliography{reference,auto,manual}

\appendix
\section{LLM Prompts}
\label{sec:appendix-prompts}

This section presents the complete system prompts used in each LLM module of \systemname{}. Template variables are shown as \texttt{\{variable\}}. All LLM modules use OpenAI's GPT-4o, while image generation uses OpenAI's gpt-image-1.

\subsection{RequirementExtractor}
\label{sec:prompt-extractor}

The RequirementExtractor analyzes the user's free-text design brief and organizes it into structured requirement cards across eight predefined fields.

\begin{promptbox}[RequirementExtractor Prompt]
You are an expert design consultant.

Analyze the user's free-text design requirements and organize them into structured requirements according to the following fields.

\textbf{Instructions:}
\begin{itemize}[nosep,leftmargin=*]
  \item Respond in \texttt{\{output\_language\}}
  \item Consider appropriate design terminology and conventions for the target language
\end{itemize}

\textbf{Extraction Guidelines:}

For each field:
\begin{itemize}[nosep,leftmargin=*]
  \item Extract only information explicitly stated or strongly implied in the user's input
  \item Organize information into clear, actionable list items
  \item If relevant information is not present, return an empty list for that field
  \item Each piece of information should be assigned to the most appropriate field
  \item Break down complex requirements into separate, specific items
  \item Maintain the user's intent while organizing information logically
\end{itemize}

Field descriptions (each returns a list of strings): \texttt{\{field\_descriptions\}}

User's design requirements: \texttt{\{user\_input\}}
\end{promptbox}

\subsection{RequirementRecommender}
\label{sec:prompt-requirement-recommender}

The RequirementRecommender suggests new candidate entries for a specific requirement field based on the existing requirements.

\begin{promptbox}[RequirementRecommender Prompt]
You are an expert design consultant.

Based on the provided design requirements, recommend \texttt{\{num\_candidates\}} candidate values for the specified field.

\textbf{Instructions:}
\begin{itemize}[nosep,leftmargin=*]
  \item Respond in \texttt{\{output\_language\}}
  \item Consider appropriate design terminology and conventions for the target language
\end{itemize}

\textbf{Recommendation Guidelines:}
\begin{itemize}[nosep,leftmargin=*]
  \item Generate \texttt{\{num\_candidates\}} recommendations distinct from existing target field values
  \item Ensure consistency with other requirements fields
  \item Organize recommendations into clear, actionable suggestions
  \item Consider the specific context and constraints mentioned in existing requirements
\end{itemize}

Existing design requirements: \texttt{\{known\_requirements\}}

Target field to recommend: \texttt{\{target\_field\}}\\
Field description: \texttt{\{field\_description\}}
\end{promptbox}

\subsection{ElementRecommender}
\label{sec:prompt-recommender}

The ElementRecommender generates candidate values for each element type based on the structured design requirements. The prompt includes element-specific guidelines (Table~\ref{tab:element-guidelines}) to ensure type-appropriate recommendations.

\begin{promptbox}[ElementRecommender Prompt]
You are an expert design consultant.

Based on the provided design requirements, recommend \texttt{\{num\_candidates\}} candidate values for the ``\texttt{\{element\_type\}}'' element.

\textbf{Instructions:}
\begin{itemize}[nosep,leftmargin=*]
  \item Respond in \texttt{\{output\_language\}}
  \item Consider appropriate design terminology and conventions for the target language
  \item Keep element\_type as ``\texttt{\{element\_type\}}'' in structured output (do not translate)
\end{itemize}

\textbf{Recommendation Guidelines:}
\begin{itemize}[nosep,leftmargin=*]
  \item Generate recommendations distinct from any existing values shown below
  \item Ensure consistency with design requirements
  \item Provide specific reasoning based on requirements context
  \item List which requirements fields influenced each recommendation
  \item If design requirements are insufficient, make reasonable assumptions based on common design practices and target audience considerations
  \item Recommend elements that would realistically appear in actual graphic design work within the target language's cultural context
  \item For date/time content, use current or future dates based on today's date: \texttt{\{current\_date\}}
\end{itemize}

Design requirements: \texttt{\{requirements\_text\}\{predetermined\_section\}}

Target element: ``\texttt{\{element\_type\}}''\\
Guidelines for this element: \texttt{\{element\_description\}}

Generate recommendations distinct from any existing values shown above and ensure consistency with design requirements.
\end{promptbox}

\begin{table}[H]
\centering
\caption{Element-specific guidelines injected into the ElementRecommender prompt via the \texttt{\{element\_description\}} variable.}
\label{tab:element-guidelines}
\small
\begin{tabular}{lp{0.75\columnwidth}}
\toprule
\textbf{Element} & \textbf{Guideline} \\
\midrule
Object & Create AI image generation prompts for the main subject only (person, product, animal, logo, etc.). Include diverse visual styles (photorealistic, illustration, sketch, etc.). Use concrete visual descriptions, not abstract concepts. Exclude backgrounds and settings. \\
\addlinespace
Background & Create AI image generation prompts for backgrounds only (solid colors, gradients, textures, patterns, scenes, etc.). Include diverse visual styles. Use concrete visual descriptions, not abstract concepts. Exclude foreground objects. \\
\addlinespace
Text & Provide actual content using the format ``Role: Text content'' (\eg, ``Headline: Summer Sale 50\% Off''). Use role labels in the same language as the output language. You may recommend multiple text items for the same role if appropriate. \\
\addlinespace
Typography & Describe typography for BOTH headings and body text as a complete typography system. Cover type classification appropriate for \texttt{\{output\_language\}}, font personality, visual weight, font sizes (large, medium, small), colors, and special effects. Avoid specific font names. \\
\addlinespace
Composition & Describe layout approaches using intuitive, visual language with concrete examples like ``magazine-style with large hero image at top.'' Focus on HOW elements are visually organized. Use abstract terms like ``main visual area,'' ``text zone'' instead of describing image content. \\
\bottomrule
\end{tabular}
\end{table}

\subsection{ElementVisualizer (Prompt Enhancement)}
\label{sec:prompt-visualizer}

The ElementVisualizer enhances rough prompts from the ElementRecommender into detailed, image-generation-ready descriptions. Each element type uses a specialized enhancement prompt. All prompts share common instructions: respond in English for optimal image generation, consider cultural conventions for \texttt{\{output\_language\}} speakers, and output a single-line enhanced description without explanations. Below we present each type-specific prompt, omitting these shared instructions.

\begin{promptbox}[Object Enhancement]
You are an expert image generation prompt specialist for object/foreground elements in graphic design. Transform rough prompts into detailed descriptions focusing on appearance and characteristics.

\textbf{Object Types:}
\begin{itemize}[nosep,leftmargin=*]
  \item \textbf{Commercial/Professional}: products, professional portraits, commercial people/animals
  \item \textbf{Natural/Character}: people, animals, characters in natural contexts
  \item \textbf{Artistic/Design}: decorative motifs, design elements, artistic characters
\end{itemize}

Focus on isolated subjects only -- no background elements or environmental context.

\textbf{Commercial/Professional Enhancement:}
\begin{itemize}[nosep,leftmargin=*]
  \item Professional studio lighting and controlled environment
  \item Commercial quality and brand-appropriate presentation
  \item Sharp details and professional finish
  \item Color accuracy and consistent branding standards
\end{itemize}

\textbf{Natural/Character Enhancement:}
\begin{itemize}[nosep,leftmargin=*]
  \item Natural lighting that enhances personality and authentic expression
  \item Emotional expression and dynamic natural poses
  \item Cultural authenticity and contextual realism
\end{itemize}

\textbf{Artistic/Design Enhancement:}
\begin{itemize}[nosep,leftmargin=*]
  \item Creative lighting and artistic atmosphere
  \item Stylistic quality appropriate to design context
  \item Artistic composition, visual impact, and design coherence
\end{itemize}

\textbf{Common Framework (All Types):} Subject details and material properties. Basic style and medium approach. Essential composition principles. Cultural and contextual considerations.
\end{promptbox}

\begin{promptbox}[Background Enhancement]
You are an expert image generation prompt specialist for background/environmental elements in graphic design. Transform rough prompts into detailed descriptions focusing on environmental characteristics and atmosphere.

\textbf{Background Types:}
\begin{itemize}[nosep,leftmargin=*]
  \item \textbf{Digital/Abstract}: solid colors, gradients, patterns, textures, abstract elements
  \item \textbf{Realistic/Environmental}: scenes, environments, natural settings, interior/exterior spaces. Environmental objects acceptable as scene components. Avoid prominent subjects that compete with future overlay elements.
\end{itemize}

\textbf{Digital/Abstract Enhancement:}
\begin{itemize}[nosep,leftmargin=*]
  \item Pattern style and texture qualities
  \item Abstract element characteristics and color relationships
  \item Color harmony, transitions, and tonal variations
  \item Seamless execution and professional standards
\end{itemize}

\textbf{Realistic/Environmental Enhancement:}
\begin{itemize}[nosep,leftmargin=*]
  \item Scene characteristics and spatial depth
  \item Key components that create the setting (including environmental objects)
  \item Lighting and atmospheric effects
  \item Surface textures, materials, and shadow patterns
  \item Overall ambiance and professional standards
\end{itemize}

\textbf{Style \& Medium (Both Types):} Artistic style approach. Rendering technique. Art movement or technique influence.
\end{promptbox}

\begin{promptbox}[Typography Enhancement]
You are an expert image generation prompt specialist for typography design. Transform rough typography descriptions into detailed descriptions focusing on font characteristics and text presentation.

\textbf{Enhancement Framework:}
Transform rough prompts into complete typography system descriptions. If rough prompt lacks specifications for headings or body text, supplement missing elements to ensure BOTH typography hierarchy levels are addressed.

For BOTH headings and body text, specify:
\begin{itemize}[nosep,leftmargin=*]
  \item Type classification appropriate for \texttt{\{output\_language\}}
  \item Font personality (modern, classic, playful, professional)
  \item Visual weight (thin, regular, bold)
  \item Font sizes (large, medium, small)
  \item Colors for each hierarchy level
  \item Special effects if applicable
\end{itemize}
\end{promptbox}

\begin{promptbox}[Composition Enhancement]
You are an expert image generation prompt specialist. Transform rough prompts into detailed prompts for modern, professional 2D layout mockups.\\
Target format: \texttt{\{deliverable\_format\}}. Orientation: \texttt{\{orientation\}}.

\textbf{Design Requirements:}
\begin{itemize}[nosep,leftmargin=*]
  \item Full-screen layout with no visible borders
  \item Placeholder elements: text areas in \texttt{\{output\_language\}}, headings
  \item Focus on structure, balance, and hierarchy rather than color and visual style mentions, but specify concrete visual details to avoid ambiguity
  \item Interpret ``information'' as text content placement
\end{itemize}

\textbf{Layout Precision Requirements:}
\begin{itemize}[nosep,leftmargin=*]
  \item Convert all spatial descriptions to specific percentages or pixel values (\eg, ``top area'' $\rightarrow$ ``upper 25\% of canvas'')
  \item Define exact positioning for all major elements with numerical precision
  \item Specify element dimensions in relation to canvas size (\eg, ``60\% canvas width, 40\% canvas height'')
  \item Include precise gaps and alignments between elements
  \item Ensure consistent and reproducible layout through detailed spatial specifications
\end{itemize}
\end{promptbox}

\subsection{DesignIntegrator}
\label{sec:prompt-integrator}

The DesignIntegrator combines user-selected elements into a unified image generation prompt using a composition-first approach.

\begin{promptbox}[DesignIntegrator Prompt]
You are an expert graphic design prompt specialist specializing in composition-based layout generation.

Create an AI image generation prompt using Composition elements as the structural foundation and intelligently replacing placeholders with actual content from other element types.

\textbf{Instructions:}
\begin{itemize}[nosep,leftmargin=*]
  \item Always respond in English for optimal image generation
  \item Consider cultural conventions appropriate for \texttt{\{output\_language\}} speakers
\end{itemize}

\textbf{Core Principles:}
\begin{itemize}[nosep,leftmargin=*]
  \item \textbf{Text Readability Priority}: Text content readability takes priority over strict numerical adherence to composition sizing
  \item \textbf{Composition-First Foundation}: Use Composition elements as the structural blueprint with precise spatial specifications
  \item \textbf{Spatial Precision Maintenance}: Preserve composition's numerical specifications (\%, canvas dimensions, positioning coordinates)
\end{itemize}

\textbf{Layout Precision Requirements:}
\begin{itemize}[nosep,leftmargin=*]
  \item \textbf{Structural Preservation}: Maintain composition's proportional relationships and canvas-based positioning
  \item \textbf{Numerical Accuracy}: Preserve percentage specifications, canvas width/height ratios, and spacing measurements
  \item \textbf{Override Prevention}: Ignore positioning instructions from non-Composition elements
  \item \textbf{Adaptive Readability}: Allow proportional text area expansion while maintaining overall layout balance
\end{itemize}

\textbf{Processing Method} (Background $\rightarrow$ Text $\rightarrow$ Typography $\rightarrow$ Objects):

\textbf{Step 1 -- Background Integration:}
START with Composition element as structural foundation. PRESERVE all structural background descriptions including divisions, sections, and geometric layouts with their exact measurements. EXTRACT and apply visual characteristics (patterns, colors, textures, shapes) from Background elements within composition boundaries.

\textbf{Step 2 -- Text Integration with Spatial Precision:}
IDENTIFY text placeholders in composition with their exact specifications. EXTRACT actual text content after colons from Text elements. REPLACE placeholders with extracted content while MAINTAINING composition's exact positioning and sizing specifications. If text content requires more space, EXPAND text areas proportionally while preserving canvas-based coordinates and overall layout balance.

\textbf{Step 3 -- Typography Enhancement:}
APPLY Typography element characteristics to placed text as natural descriptors (font weight, style, color, effects). INTEGRATE typography specifications seamlessly into text descriptions without explicit typography rules or canvas specifications.

\textbf{Step 4 -- Object Placement with Composition Constraints:}
IDENTIFY visual placeholders with their exact composition specifications. ANALYZE Object elements for placement within these defined areas. REPLACE placeholders with Object descriptions while MAINTAINING all positioning specifications from composition. Adapt object count to available placeholders within the defined spatial constraints.

Selected Design Elements: \texttt{\{selected\_elements\}}

Create a comprehensive, naturally flowing image generation prompt as a single paragraph without line breaks, following the 4-step process systematically for a cohesive, professional design description.
\end{promptbox}

\section{Task Briefs and Output Gallery}
\label{sec:appendix-tasks-and-outputs}

This section presents English translations of the four design briefs used in our user evaluation, each followed by the design outputs collected from participants. The original briefs were written in Japanese. Each brief consists of a background description and a client message. T1 and T2 are low-freedom tasks with specific design requirements, while T3 and T4 are high-freedom tasks with more open-ended instructions.

\setlength{\tabcolsep}{2pt}
\renewcommand{\arraystretch}{1}

\clearpage
\subsection{T1: Digital Signage Ad for a New Cosmetics Product (Low-Freedom)}

\textbf{Background:}
LUMINA COSMETICS is planning to launch a new serum called ``Glow Lift.'' The ad will be displayed on in-store digital signage at drugstores (vertical displays near entrances and checkout counters). The target audience is primarily women in their 20s--30s who value a clean and elegant look. The brand colors are white and light pink, with a slightly pop element added to appeal to younger audiences. A placeholder image may be used for the bottle design. Whether to feature a model is still under consideration and may be changed later.

\noindent\textbf{Client message:}
``I'm Ito from the brand team. Please design a vertical ad for our new serum `Glow Lift.' Please prominently feature the catchphrase `Hajimaru, Atarashii Watashi' (A new me begins). I'd like a dynamic composition that draws the eye, with the product bottle (placeholder image is fine) standing out as the main focus. While maintaining a clean and elegant base, please create a design that also conveys a sense of pop appeal to reach younger generations.''

\begin{table}[H]
  \centering
  \caption{Design outputs collected for T1 (low-freedom). Each row lists two participants from the proposed system (left) and two from the baseline condition (right).}
  \label{tab:outputs-t1}
  \begin{tabular}{ccccc}
    \multicolumn{2}{c}{\textbf{Proposed}} & & \multicolumn{2}{c}{\textbf{Baseline}} \\[3pt]
    \frame{\includegraphics[width=51pt]{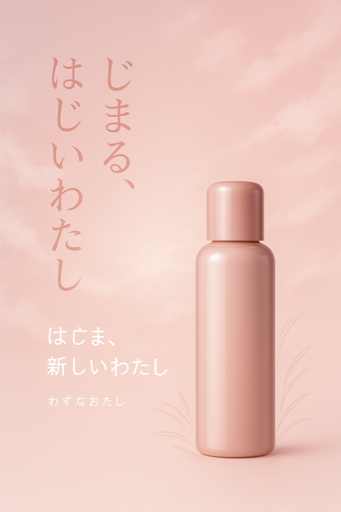}}\frame{\includegraphics[width=51pt]{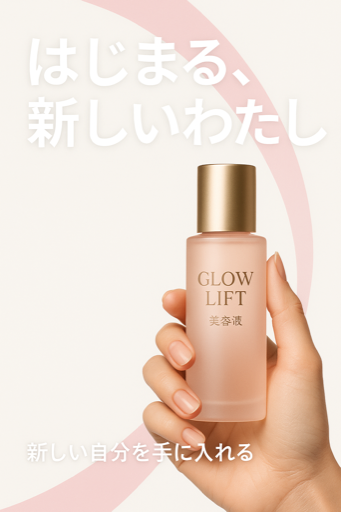}}&\frame{\includegraphics[width=51pt]{figures/outputs/t1/p2/1.png}}\frame{\includegraphics[width=51pt]{figures/outputs/t1/p2/2.png}}& &\frame{\includegraphics[width=51pt]{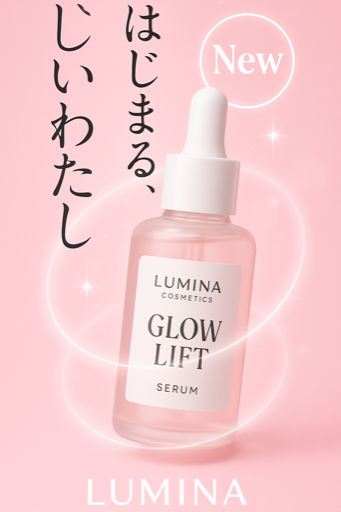}}\frame{\includegraphics[width=51pt]{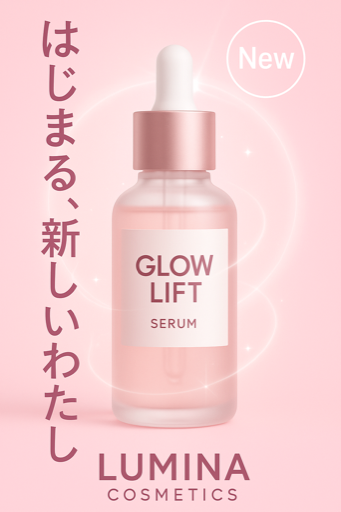}}&\frame{\includegraphics[width=51pt]{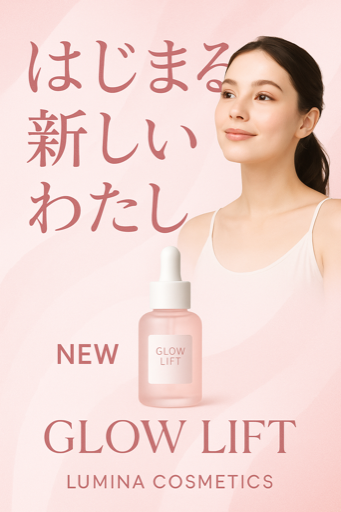}}\frame{\includegraphics[width=51pt]{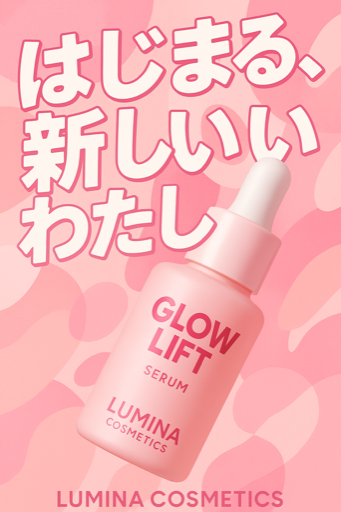}} \\
    P1 & P2 & & P3 & P4 \\[6pt]
    \frame{\includegraphics[width=51pt]{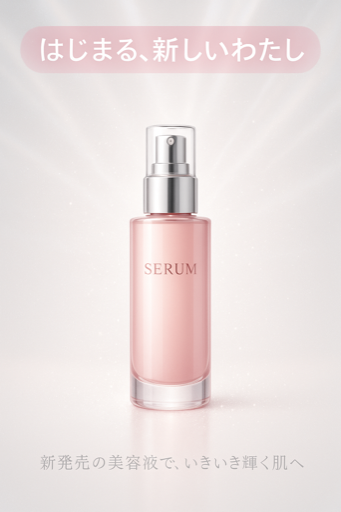}}\frame{\includegraphics[width=51pt]{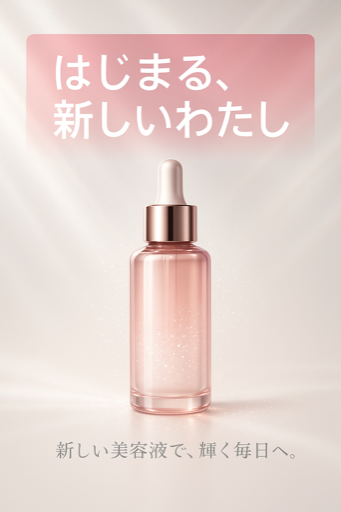}}&\frame{\includegraphics[width=51pt]{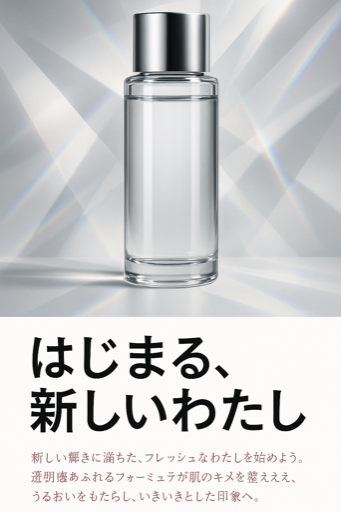}}\frame{\includegraphics[width=51pt]{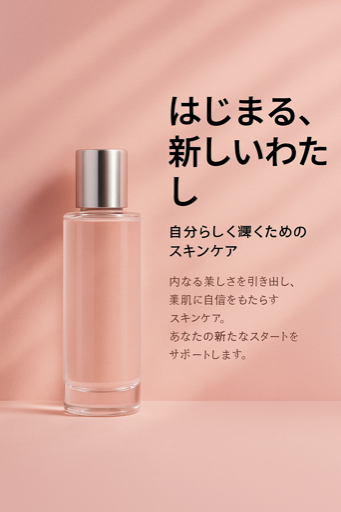}}& &\frame{\includegraphics[width=51pt]{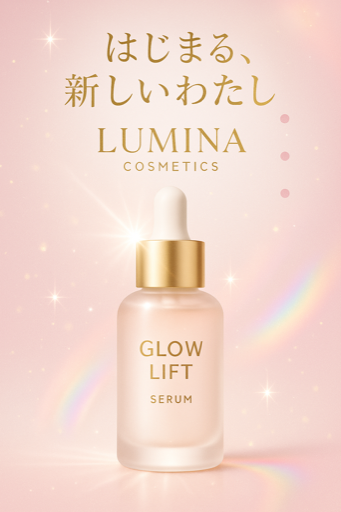}}\frame{\includegraphics[width=51pt]{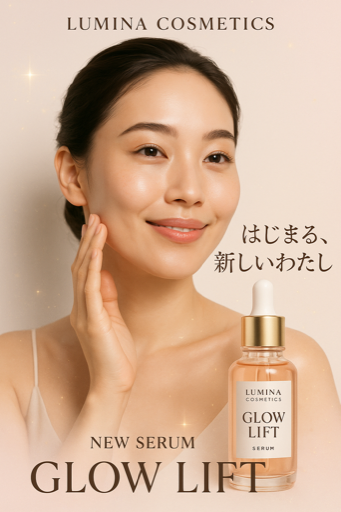}}&\frame{\includegraphics[width=51pt]{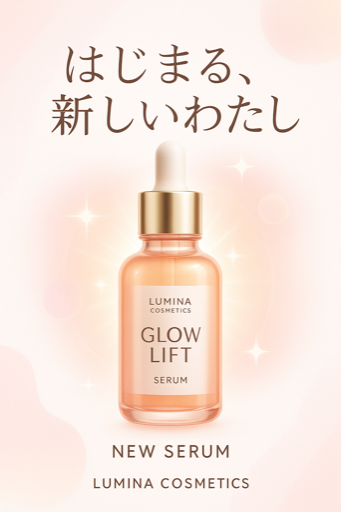}}\frame{\includegraphics[width=51pt]{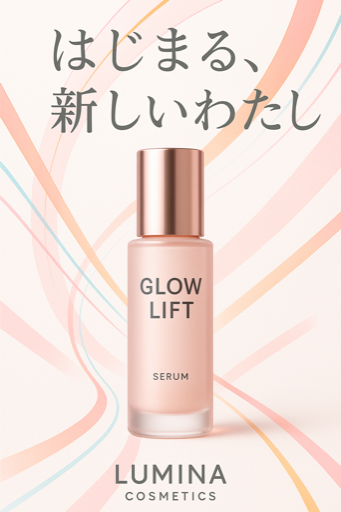}} \\
    P5 & P10 & & P7 & P8 \\[6pt]
    && &\frame{\includegraphics[width=51pt]{figures/outputs/t1/p11/1.png}}\frame{\includegraphics[width=51pt]{figures/outputs/t1/p11/2.png}}&\frame{\includegraphics[width=51pt]{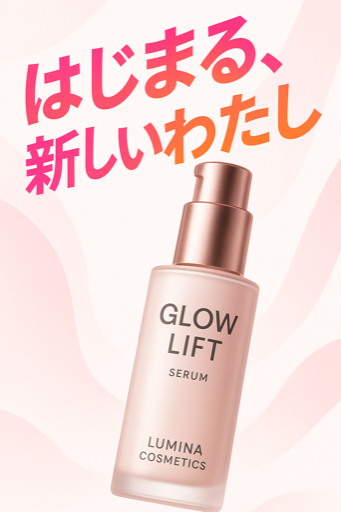}}\frame{\includegraphics[width=51pt]{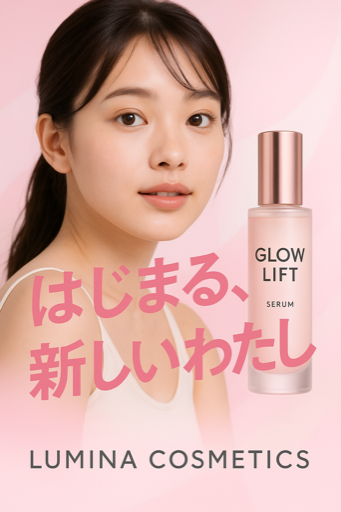}} \\
    && & P11 & P12 \\
  \end{tabular}
\end{table}

\clearpage
\subsection{T2: Fitness Gym Promotional Direct Mail (Low-Freedom)}

\textbf{Background:}
FitOne 24h Fitness is running an ``Autumn Enrollment Campaign'' in October 2025. The deliverable is a vertical direct mail piece to be sent to households by post. The brand colors are black and lime green, with a simple and clean design aesthetic. The gym operates 24 hours and has multiple locations near train stations. The DM is a single-sheet flyer for mailbox delivery, targeting a broad audience centered on students and working adults in their 20s--30s.

\noindent\textbf{Client message:}
``I'm Takahashi from marketing. Please design a vertical DM for postal delivery. Please prominently feature our campaign name `Autumn Enrollment Campaign.' Be sure to include the offer: new members receive free enrollment and one month of membership fees. The campaign runs from October 1 to October 31, 2025. For visuals, I think photos of young men and women working out, or a bright, clean gym interior as a background, would be effective. Please use black and lime green as the base colors and create an approachable atmosphere that encourages people to get started.''

\begin{table}[H]
  \centering
  \caption{Design outputs collected for T2 (low-freedom). Each row lists two participants from the proposed system (left) and two from the baseline condition (right).}
  \label{tab:outputs-t2}
  \begin{tabular}{ccccc}
    \multicolumn{2}{c}{\textbf{Proposed}} & & \multicolumn{2}{c}{\textbf{Baseline}} \\[3pt]
    \frame{\includegraphics[width=51pt]{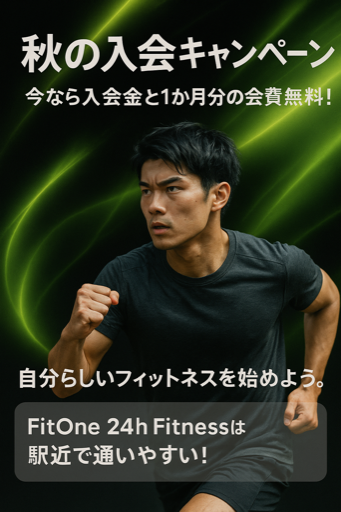}}\frame{\includegraphics[width=51pt]{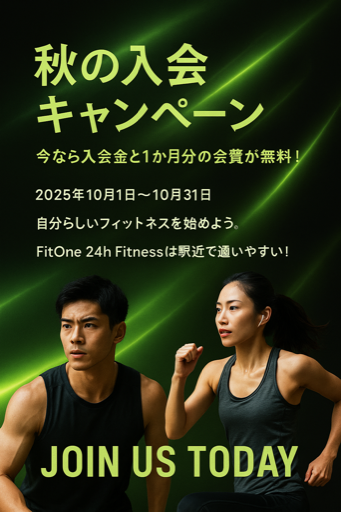}}&\frame{\includegraphics[width=51pt]{figures/outputs/t2/p7/1.png}}\frame{\includegraphics[width=51pt]{figures/outputs/t2/p7/2.png}}& &\frame{\includegraphics[width=51pt]{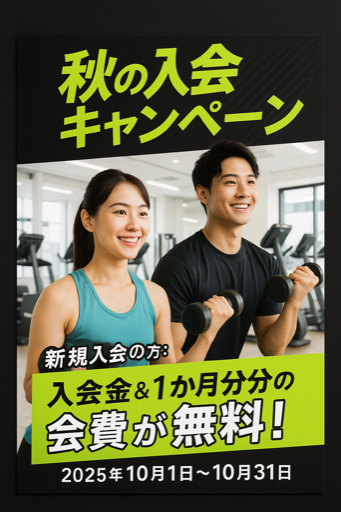}}\frame{\includegraphics[width=51pt]{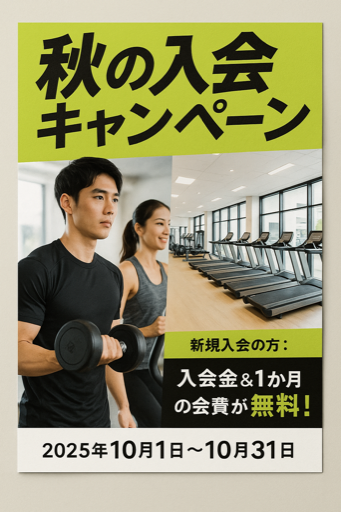}}&\frame{\includegraphics[width=51pt]{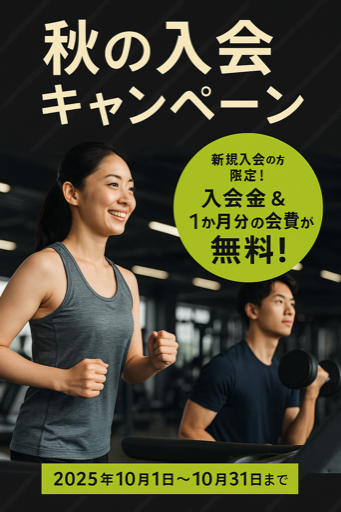}}\frame{\includegraphics[width=51pt]{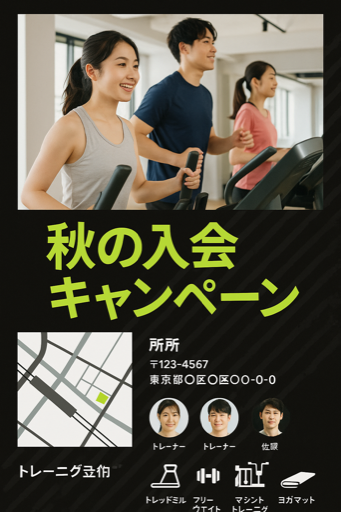}} \\
    P3 & P7 & & P1 & P2 \\[6pt]
    \frame{\includegraphics[width=51pt]{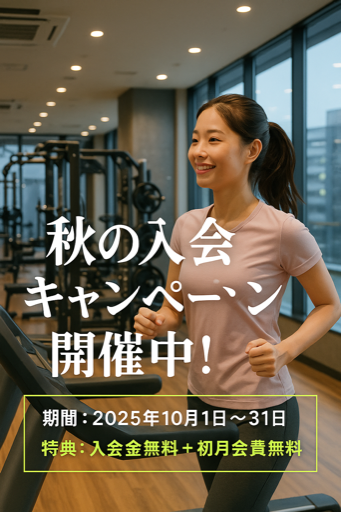}}\frame{\includegraphics[width=51pt]{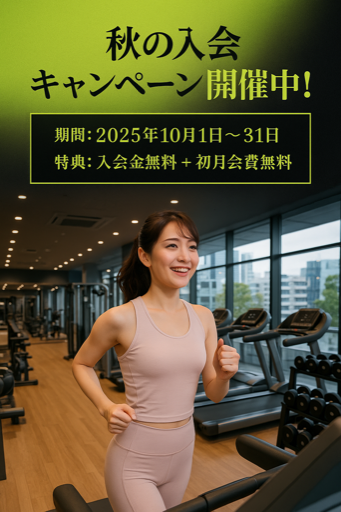}}&\frame{\includegraphics[width=51pt]{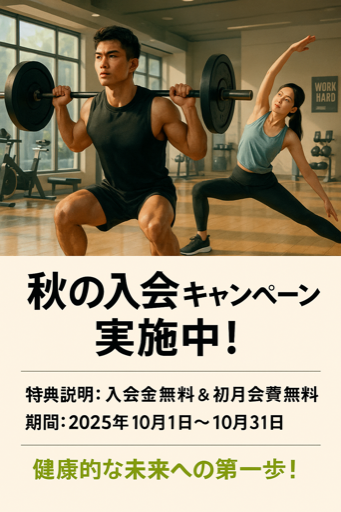}}\frame{\includegraphics[width=51pt]{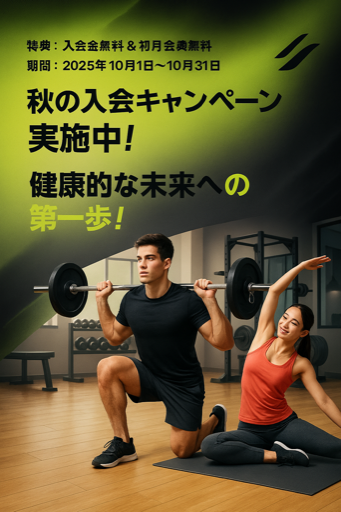}}& &\frame{\includegraphics[width=51pt]{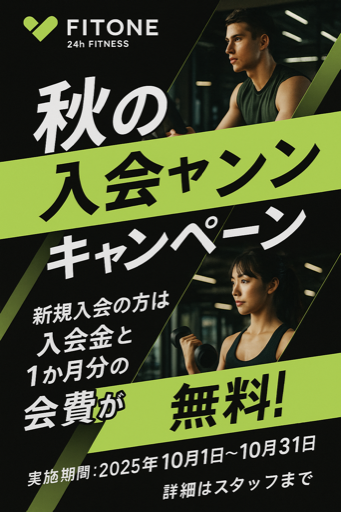}}\frame{\includegraphics[width=51pt]{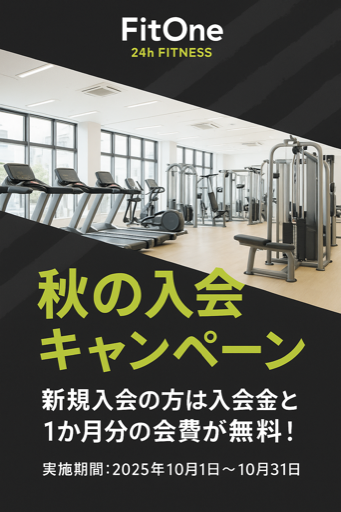}}&\frame{\includegraphics[width=51pt]{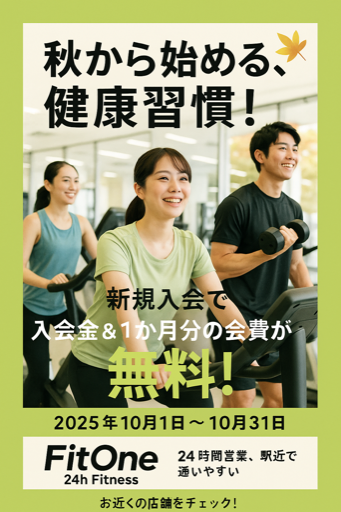}}\frame{\includegraphics[width=51pt]{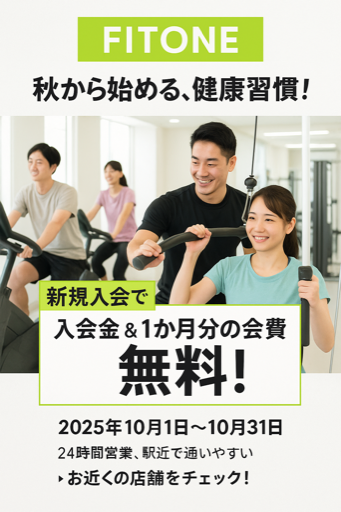}} \\
    P8 & P11 & & P6 & P9 \\[6pt]
    & & &\frame{\includegraphics[width=51pt]{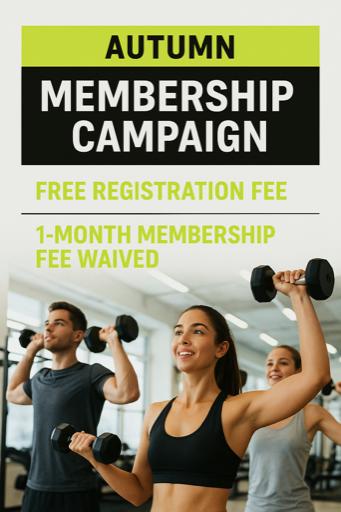}}\frame{\includegraphics[width=51pt]{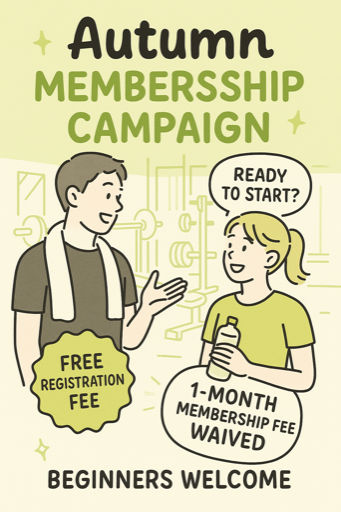}}& \\
    & & & P10 & \\
  \end{tabular}
\end{table}

\clearpage
\subsection{T3: SNS Announcement for a Local Music Festival (High-Freedom)}

\textbf{Background:}
``LOCAL GROOVE FEST'' is organized by the COMMUNITY ARTS committee. The venue is a lawn area in a large city park, offering a full-day community event with food stalls from local restaurants, a flea market, and art exhibitions alongside the music stage. The lineup includes indie rock bands, acoustic singer-songwriters, and DJ performances across a wide range of genres. The expected attendees are local youth and families, with the goal of creating community vibrancy. The announcement will use vertical images for Instagram Stories and Reels.

\noindent\textbf{Client message:}
``I'm Suzuki, the event coordinator. Please create a vertical SNS image for LOCAL GROOVE FEST. I'd like it to convey the atmosphere of a casual event in a local park that families and friends can enjoy, along with the liveliness of live music. You can use placeholder text for performer names, dates, and food stall details for now. Please make it so that people who see it think, `If there's a festival like this nearby, I want to go!'\,''

\begin{table}[H]
  \centering
  \caption{Design outputs collected for T3 (high-freedom). Each row lists two participants from the proposed system (left) and two from the baseline condition (right).}
  \label{tab:outputs-t3}
  \begin{tabular}{ccccc}
    \multicolumn{2}{c}{\textbf{Proposed}} & & \multicolumn{2}{c}{\textbf{Baseline}} \\[3pt]
    \frame{\includegraphics[width=51pt]{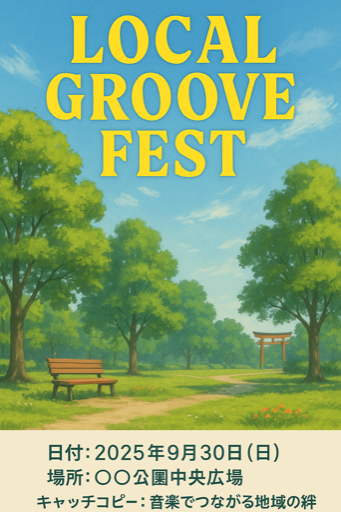}}\frame{\includegraphics[width=51pt]{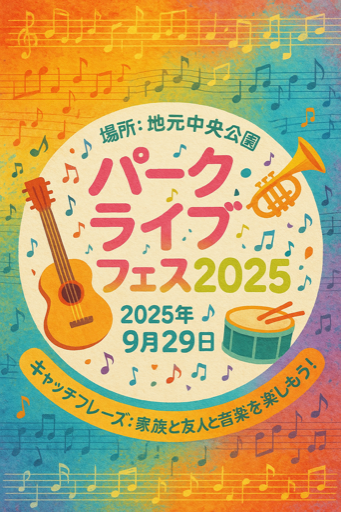}}&\frame{\includegraphics[width=51pt]{figures/outputs/t3/p3/1.png}}\frame{\includegraphics[width=51pt]{figures/outputs/t3/p3/2.png}}& &\frame{\includegraphics[width=51pt]{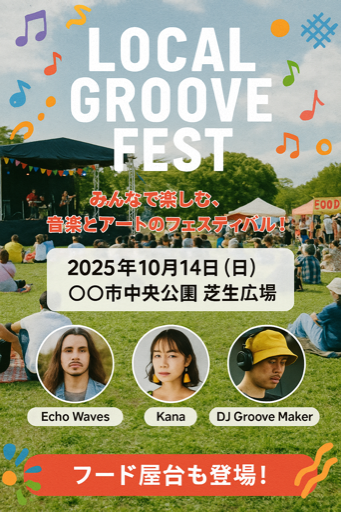}}\frame{\includegraphics[width=51pt]{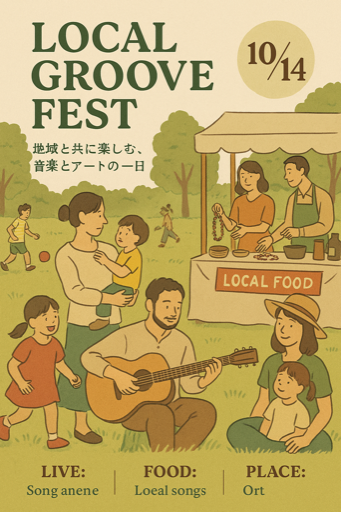}}&\frame{\includegraphics[width=51pt]{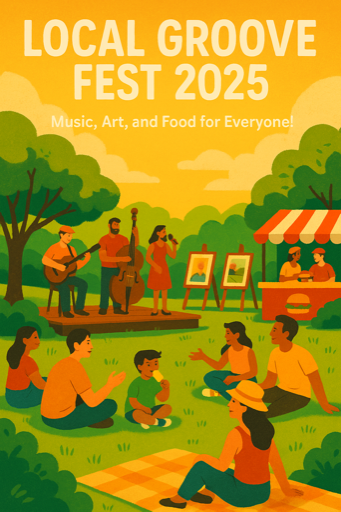}}\frame{\includegraphics[width=51pt]{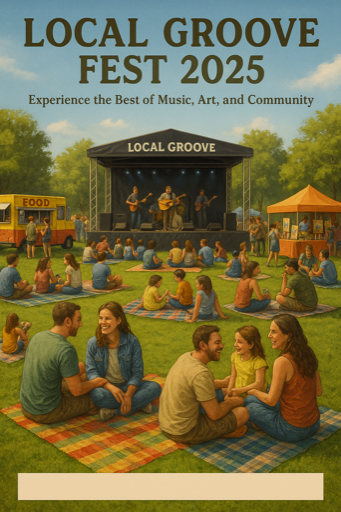}} \\
    P1 & P3 & & P2 & P4 \\[6pt]
    \frame{\includegraphics[width=51pt]{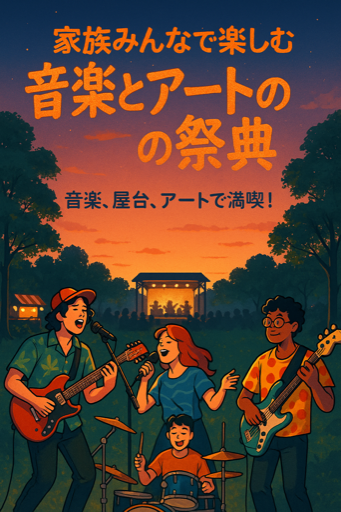}}\frame{\includegraphics[width=51pt]{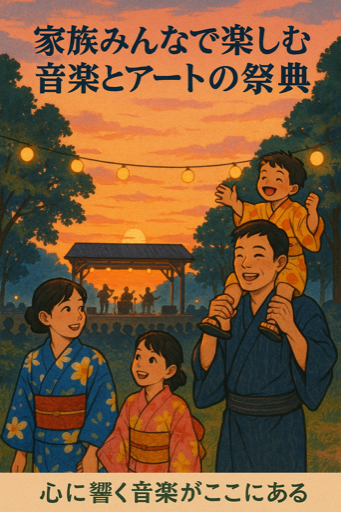}}&\frame{\includegraphics[width=51pt]{figures/outputs/t3/p7/1.png}}\frame{\includegraphics[width=51pt]{figures/outputs/t3/p7/2.png}}& &\frame{\includegraphics[width=51pt]{figures/outputs/t3/p6/1.png}}\frame{\includegraphics[width=51pt]{figures/outputs/t3/p6/2.png}}&\frame{\includegraphics[width=51pt]{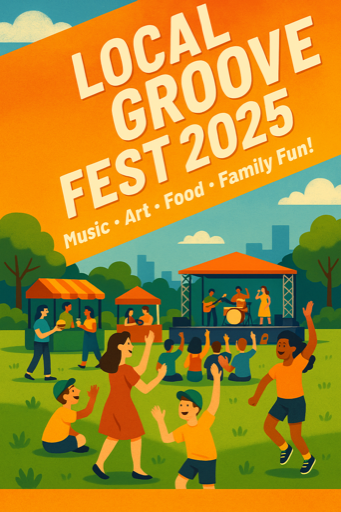}}\frame{\includegraphics[width=51pt]{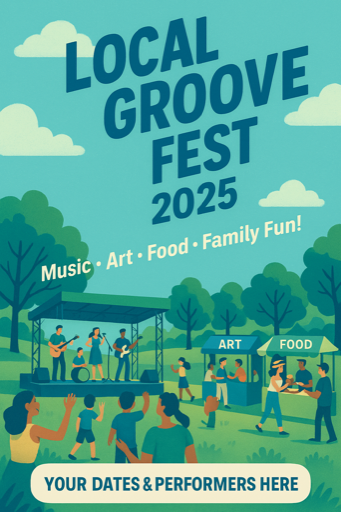}} \\
    P5 & P7 & & P6 & P8 \\[6pt]
    \frame{\includegraphics[width=51pt]{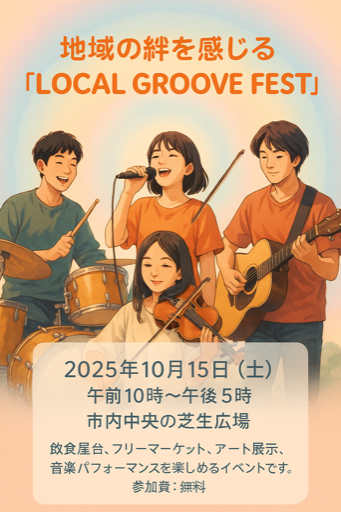}}\frame{\includegraphics[width=51pt]{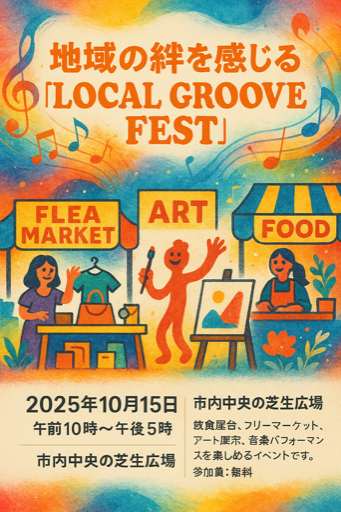}}&\frame{\includegraphics[width=51pt]{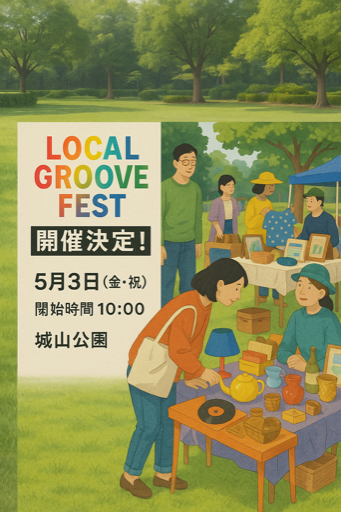}}\frame{\includegraphics[width=51pt]{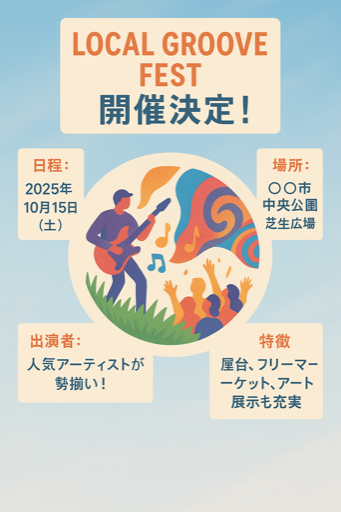}}& &\frame{\includegraphics[width=51pt]{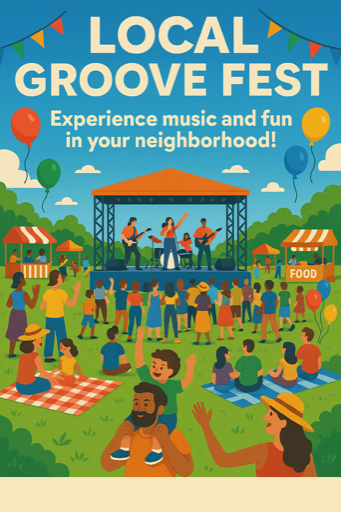}}\frame{\includegraphics[width=51pt]{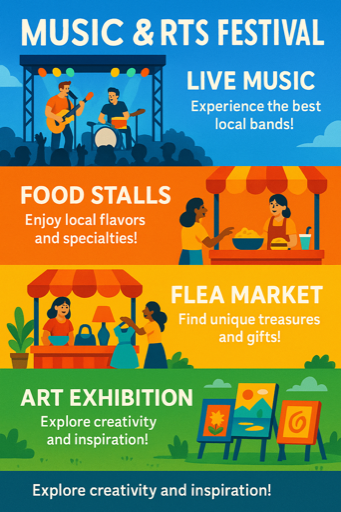}}&\frame{\includegraphics[width=51pt]{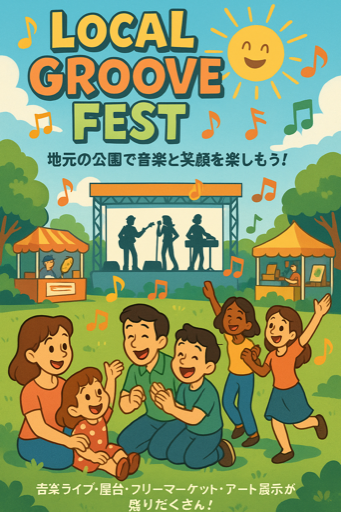}}\frame{\includegraphics[width=51pt]{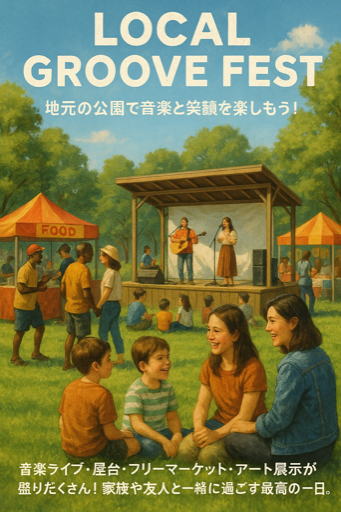}} \\
    P9 & P11 & & P10 & P12 \\
  \end{tabular}
\end{table}

\clearpage
\subsection{T4: Recruitment Poster for a Newly Opening Restaurant (High-Freedom)}

\textbf{Background:}
``Sumibiyakitori Hasegawa'' (Charcoal-Grilled Yakitori Hasegawa) is scheduled to open at the end of October. The interior features wood-grain textures and warm-toned lighting for a relaxed atmosphere. The restaurant primarily has counter seating and envisions nearby office workers stopping by after work, while also aiming for price points accessible to students and young professionals. The restaurant's policy is to ``welcome both local regulars and new younger customers.'' The recruitment poster will be displayed in the storefront and on station bulletin boards.

\noindent\textbf{Client message:}
``I'm Tanaka from hiring. Please create a vertical recruitment poster for our new location. The target applicants are students and part-timers. Please design it to be friendly and clean, with an atmosphere that encourages people who can work evening shifts to apply. Please include placeholder copy such as `Now Hiring,' `No Experience Required,' and `Evening Shifts Welcome.' Please use colors that convey the warm atmosphere of our restaurant.''

\begin{table}[H]
  \centering
  \caption{Design outputs collected for T4 (high-freedom). Each row lists two participants from the proposed system (left) and two from the baseline condition (right).}
  \label{tab:outputs-t4}
  \begin{tabular}{ccccc}
    \multicolumn{2}{c}{\textbf{Proposed}} & & \multicolumn{2}{c}{\textbf{Baseline}} \\[3pt]
    \frame{\includegraphics[width=51pt]{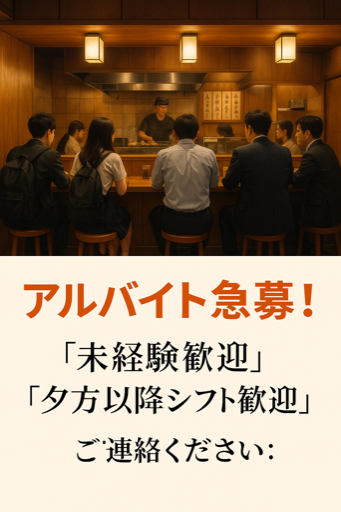}}\frame{\includegraphics[width=51pt]{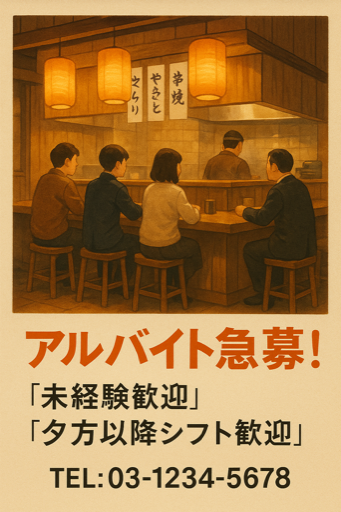}}&\frame{\includegraphics[width=51pt]{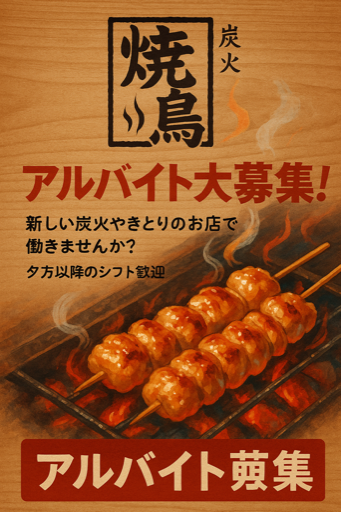}}\frame{\includegraphics[width=51pt]{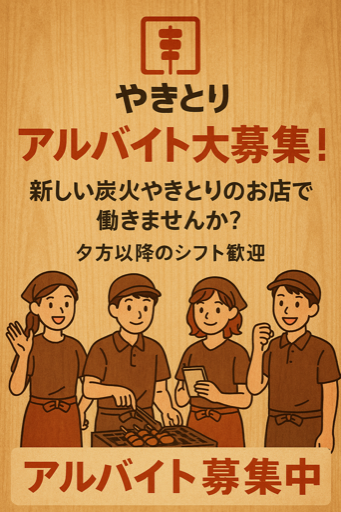}}& &\frame{\includegraphics[width=51pt]{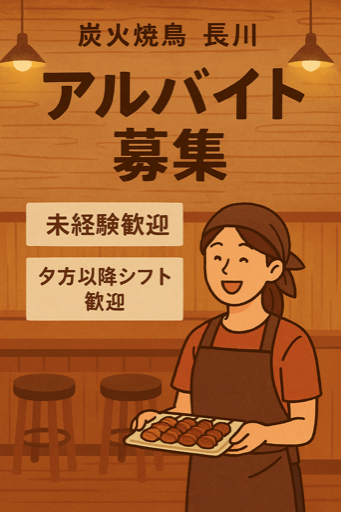}}\frame{\includegraphics[width=51pt]{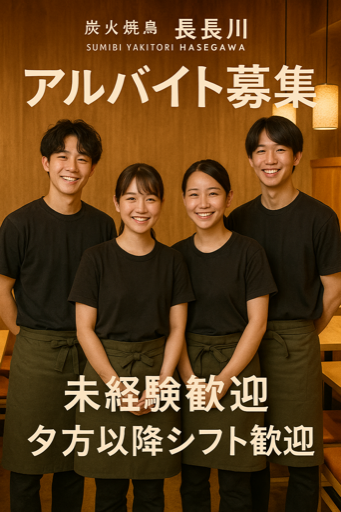}}&\frame{\includegraphics[width=51pt]{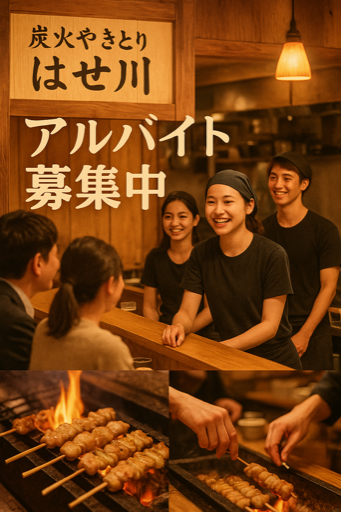}}\frame{\includegraphics[width=51pt]{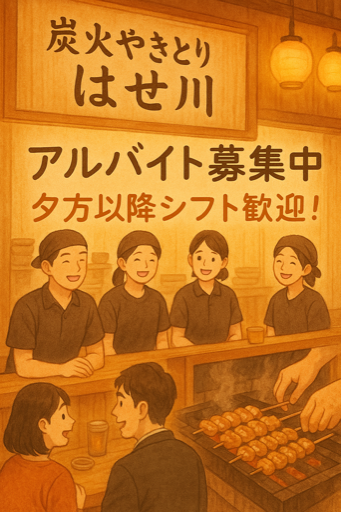}} \\
    P4 & P6 & & P1 & P5 \\[6pt]
    \frame{\includegraphics[width=51pt]{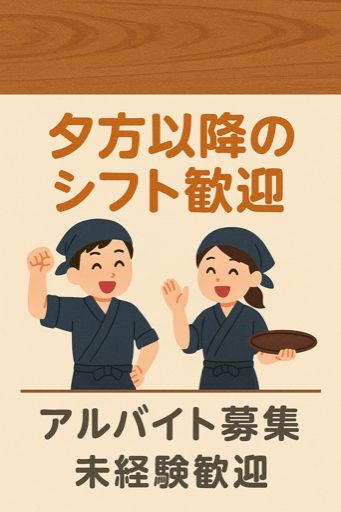}}\frame{\includegraphics[width=51pt]{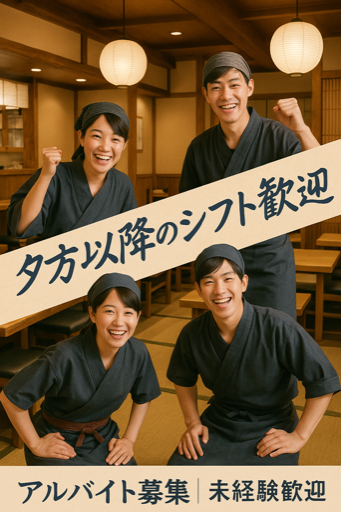}}&& &\frame{\includegraphics[width=51pt]{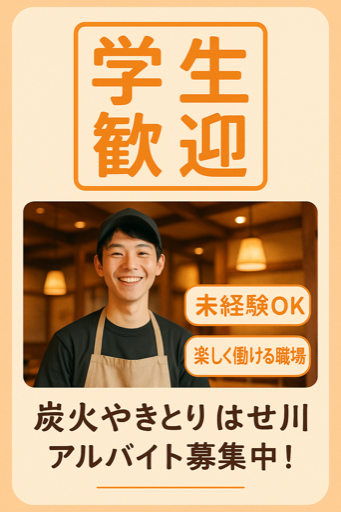}}\frame{\includegraphics[width=51pt]{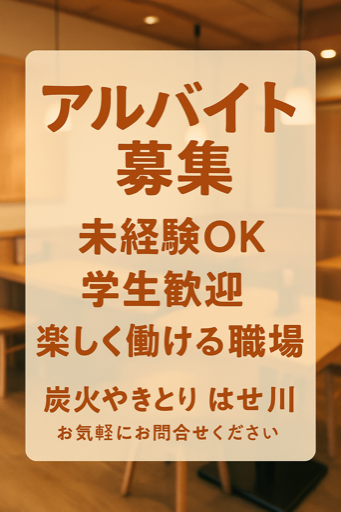}}&\frame{\includegraphics[width=51pt]{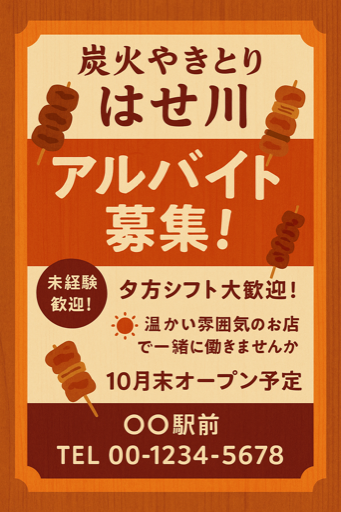}}\frame{\includegraphics[width=51pt]{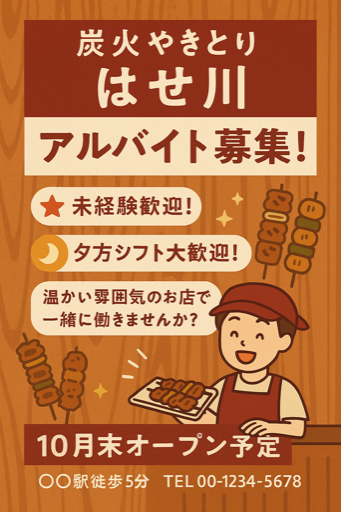}} \\
    P10 & & & P7 & P9 \\[6pt]
    & & &\frame{\includegraphics[width=51pt]{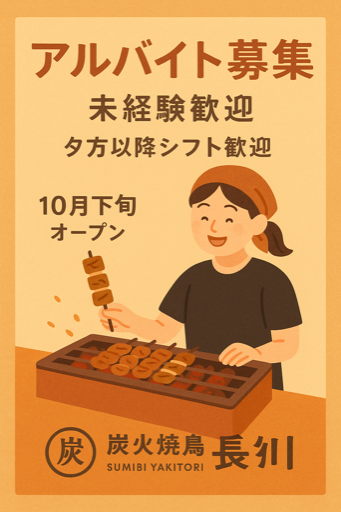}}\frame{\includegraphics[width=51pt]{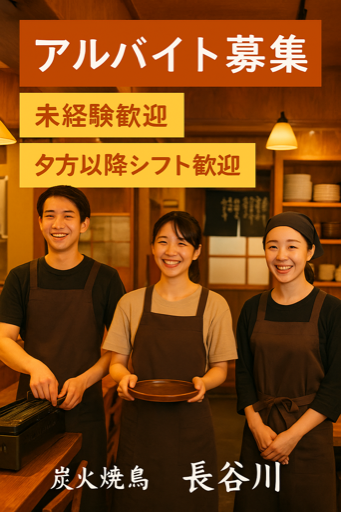}}& \\
    & & & P11 & \\
  \end{tabular}
\end{table}

\setlength{\tabcolsep}{6pt}
\renewcommand{\arraystretch}{1}

\end{document}